\documentclass[twocolumn,tighten]{aastex6}

\slugcomment{The Astrophysical Journal, in press}
\shorttitle{FUV Spectroscopy of U Gem}  
\shortauthors{Godon et al.}
\watermark{ArXiv/Astroph Version}
\setwatermarkfontsize{1.0in}

\begin{document}

\title{{\bf 
{\it HST}/COS Far Ultraviolet  Spectroscopic Analysis of 
U Geminorum Following a Wide Outburst}
\footnote{Based on observations made with the NASA/ESA {\it Hubble
Space Telescope}, obtained at the Space Telescope Science Institute, 
which is operated by AURA, Inc., under NASA contract NAS 5-26555.}  
}

\author{Patrick Godon\altaffilmark{1,2}}
\author{Michael M. Shara\altaffilmark{3}}
\author{Edward M. Sion\altaffilmark{1}} 
\author{David Zurek\altaffilmark{3}} 

\altaffiltext{1}{Department of Astrophysics \& Planetary Science, 
Villanova University,  
800 Lancaster Avenue, Villanova, PA 19085, USA}

\altaffiltext{2}{The Johns Hopkins University,
Henry A. Rowland Department of Physics and Astronomy, 
Baltimore, MD 21218, USA} 

\altaffiltext{3}{Department of Astrophysics, American Museum of 
Natural History,   
Central Park, West and 79th Street, New York, NY 10024-5192, USA}            

\email{patrick.godon@villanova.edu} 
\email{mshara@amnh.org}
\email{edward.sion@villanova.edu} 
\email{dzurek@amnh.org }

\begin{abstract}

We have used the {\it Cosmic Origins Spectrograph} (COS) on the 
{\it Hubble Space Telescope} ({\it HST}) to obtain a series of 4
far ultraviolet (FUV, 915-2148~\AA) spectroscopic observations 
of the prototypical dwarf nova U Geminorum during its cooling  
following a two-week outburst. 
Our FUV spectral analysis of the data indicates that
the white dwarf (WD) cools from a temperature of $\sim 41,500~$K, 
15 days after the peak of the outburst, to $\sim 36,250~$K, 
56 days after the peak of the outburst, assuming a massive WD
($\log(g)=8.8$) and a distance of $100.4 \pm 3.7$ pc. 
These results are self-consistent with a $\sim 1.1 M_{\odot}$ WD    
with  a $5,000\pm 200$~km radius.      
The spectra show absorption lines of H\,{\sc i}, He\,{\sc ii}, 
C\,{\sc ii iii iv}, N\,{\sc iii iv}, O\,{\sc vi}, S\,{\sc iv}, 
Si\,{\sc ii iii iv}, Al\,{\sc iii}, Ar\,{\sc iii}, and Fe\,{\sc ii}, 
but no emission features.  
We find supra-solar abundances of nitrogen confirming 
the anomalous high N/C ratio.  
The FUV lightcurve reveals a $\pm$5\% modulation with the orbital phase, 
showing dips near phase 0.25 and $\sim$0.75, where 
the spectra exhibit an increase in the depth of some 
absorption lines and in particular  
strong absorption lines from Si\,{\sc ii}, Al\,{\sc iii}, and
Ar\,{\sc iii}. The phase dependence we observe is consistent with 
material overflowing the disk rim at the hot spot, reaching a 
maximum elevation near phase 0.75, falling back at smaller radii
near phase 0.5 where it bounces off the disk surface 
and again rising above the disk near phase $\sim$0.25. 
There is a large scatter in the absorption lines' velocities, 
especially for the silicon lines, while the carbon lines seem to 
match  more closely the orbital velocity of the WD. 
This indicates that many absorption lines are affected by-  
or form in- the overflowing stream
material veiling the WD, making the analysis of the
WD spectra more difficult. 
  
\end{abstract}

\keywords{
--- novae, cataclysmic variables
--- stars: white dwarfs 
--- stars:individual (U Geminorium)  
}

\section{{\bf 
Introduction: U Gem - The Prototypical Dwarf Nova } } \label{sec:intro}

Cataclysmic Variables (CVs) are semi-detached interacting binaries
in which a white dwarf (WD) accretes matter from a Roche lobe-filling late 
type companion. When the WD has a weak or negligible 
magnetic field, the matter is accreted by means of a disk 
in which it dissipates its gravitational potential energy.   
The most common class of CVs are the dwarf novae (DNe), found mostly in a 
state of low accretion, when the disk is dim and 
the WD dominates in the 
ultraviolet (UV) band. Every few weeks to months, 
DNe undergo semi-regular outbursts (lasting days to weeks) 
in which the system     
brightens by 3-5 visual magnitudes, as the mass transfer
rate increases  due to a thermal instability in the disk. 
During outbursts, the accretion disk totally dominates  the systems 
in the UV and optical bands. 

In December 1855 the dwarf novae U Gem became  the first CV to be discovered 
\citep{hin56},  and as such it also became 
the prototypical ``UG'' 
sub-type of dwarf novae (i.e. DNe with period $P> 3$ hrs). 
For that reason, it has been extensively studied, observed
numerous time in the 
X-ray \citep[e.g.][]{mas88,szk96}, 
extreme ultraviolet \citep[e.g.][]{lon96}, 
ultraviolet \citep[e.g.][]{lon93}, 
optical \citep[e.g.][]{zha87,und06,ech07}, 
and infrared \citep[e.g.][]{pan82}, and
it has even been modeled using three-dimensional hydrodynamics
\citep[e.g.][]{kun01} 
as well as within the context of the restricted three-body 
problem \citep[e.g.][]{sma01}. 
Consequently, we restrict ourselves here below to 
summarizing only the characteristics of the
system that are directly relevant to the present work with a limited
number of references (as the complete list is extensive).

\begin{deluxetable*}{lll}[t!] 
\tablewidth{0pt}
\tablecaption{System Parameters}
\tablehead{
Parameter         & U Gem        & References         \\ 
}
\startdata
$M_1 <M_{\odot}>$ & 1.1-1.2      & \citet{sio98,lon99} \\ 
$M_2 <M_{\odot}>$ & 0.41-0.42    & \citet{lon99,sma01,ech07} \\ 
$q=M_2/M_1$       & 0.35-0.362   & \citet{lon99,sma01,ech07}  \\ 
$i   <$deg$>$     & 67-72        & \citet{zha87,lon99,und06}  \\ 
$P   <$hr$>$      & 4.24574858   & \citet{mar90,ech07,dai09}  \\ 
$m_v$             & 8.2-13.9     & \citet{szk84,rit03}  \\ 
E(B-V)            & $0.04\pm0.01$& \citet{ver87,lad91,bru94}   \\ 
$d   <$pc$>$    & $100.4\pm 3.7$ & \citet{har04}   \\
\hline            
\enddata 
\end{deluxetable*}

U Gem has a period of 0.1769061911 day 
\citep[$\sim4$ hr 15 min,][]{mar90,ech07,dai09}), 
and a mass ratio $q=0.35 \pm 0.05$ \citep{lon99,ech07} with a WD mass 
$M_{\rm wd} \sim 1.1-1.2 M_{\odot}$ \citep{sio98,lon99}.     
It has distance of $100.4 \pm 3.7$ pc, that has been determined
astrometrically by \citet{har04}, based on a re-analysis of the 
Hubble Space Telescope Fine Guidance sensor parallaxes.  
The system is a partially eclipsing binary \citep{kra62,krz65} in which 
the disk undergoes partial eclipses (around phase $0.0 \pm 0.1$)
by the donor, 
the hot spot undergoes full eclipses (just after phase 0.0)
by the donor, 
while the WD is never eclipsed \citep{sma71,war71,mar90,ech07}. 
For this to happen with a mass ratio $q\sim 0.35$,  
the inclination has to be in the range $62^{\circ}-74^{\circ}$ 
\citep{lon99,lon06}, and the value derived in the litterature is  
between $i=67^{\circ}$ and $i=72^{\circ}$ \citep{lon99,lon06,sma01,und06}, 
confirming the early estimate by \citet{zha87} of 
$i=69.7^{\circ} \pm 0.7^{\circ}$.  
The system parameters are listed in Table 1 together with their references.

U Gem goes into outburst every $\sim$118 days, 
brightening from $m_v=14.9$ 
to $m_v=8.2$ for an average outburst time of $\approx 12$ days 
\citep{szk84,rit03}. However, the system actually exhibits 
two kinds of outbursts: narrow outbursts lasting
about a week ($\sim 4-7$ days), and wide outbursts lasting about two 
weeks \citep[$\sim 14$ days,][]{sio98}. 
Ultraviolet spectroscopic observations obtained during a wide outburst 
are consistent with a steady state disk with a mass accretion rate
ranging from $\dot{M}=2 \times 10^{-9}M_{\odot}$/yr \citep{sio97}  
to $\dot{M}=7-8 \times 10^{-9}M_{\odot}$/yr \citep{pan84,fro01}.  
During the decline from outburst the mass accretion rates decreases  
\citep[$\dot{M}=6 \times 10^{-10}M_{\odot}$/yr, see][]{sio97}, 
and eventually the metal-enriched hot WD dominates the spectrum. 
A second component appears  
on the WD, possibly an optically thick hot boundary layer or 
an ``accretion belt'' \citep{lon93}, improving 
the model fit in the shorter wavelengths of 
{\it Far Ultraviolet Spectroscopic Explorer} 
({\it FUSE}) spectra of U Gem \citep[$< 960$\AA\ ,  ][]{fro01}.    

In quiescence, shortly after outburst, the WD has a temperature
still above 40,000~K \citep{lon06}, 
and cools to $\sim 30,000$~K within several months
\citep{kip91,lon94,lon95}. The length of the quiescence 
dictates the post-outburst temperature reached, which 
has occasionally been observed to be anomalously low \citep{sio98}
following a very long quiescence period. 
The exact WD temperature derived from the data depends also on the 
WD effective surface gravity $log(g)$ assumed in the modeling.  

Nonsolar metal abundances attributed to the WD photosphere 
have been found in the studies of 
U Gem \citep{sio98,lon99,fro01}, 
revealing evidence of CNO processing, namely sub-solar carbon
and silicon abundances and supra-solar nitrogen abundance.
Sub-solar aluminum has also been recorded \citep{lon99}. 
However, phase-dependent absorption has been shown to be responsible 
for the time-variable absorption seen in the FUV,   
which complicates the analysis of the WD spectra \citep{lon06}.      

While U Gem was observed previously with {\it HST} in the UV band with 
the {\it HST}/Goddard High-Resolution Spectrograph  
\citep[GHRS,][]{sio98,lon99},   
we obtained {\it HST}/COS FUV observations of U Gem extending down the 
Lyman limit (915-2148~\AA). The only previous observations covering
the Lyman region were made with the {\it Hopkins Ultraviolet Telescope (HUT)}
\citep{lon93} and {\it FUSE} \citep{fro01,lon06}. 

We present here the FUV spectral analysis of 4 {\it HST}
COS spectra obtained following a two-week outburst. 
The data obtained at the first epoch were purposedly collected 
to span a complete binary orbit ($\sim 4$ hours), 
while the data obtained at the remaining
three epochs were each collected for only a quarter of an orbit. 
In this manner we were able to analyze the data to detect any
variation as a function of the orbital phase as well as a 
function of the time passed since outburst. 
In the next section we describe the observations and data,
in section 3 we present and discuss the results from our spectral 
analysis, we then conclude in section 4 with a summary.

\section{{\bf The {\it HST}/COS Spectra}}  

\subsection{{\bf Observations}} 

\begin{figure}[h!] 
\vspace{-4.5cm}
\plotone{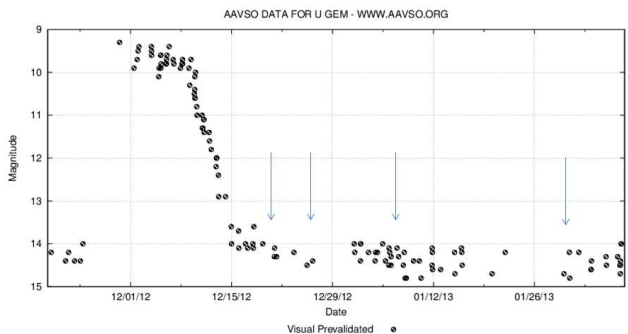}       
\caption{AAVSO optical light curve (visual magnitude vs. time) 
of U Gem at the time of COS observations.  
The four {\it HST}/COS observations were taken following an outburst lasting
15 days (``wide'' outburst), at the dates indicated with vertical arrows. 
\label{aavso} 
} 
\end{figure}

U Gem went into an outburst on November 29, 2013, for about 15 days, 
increasing its brightness from $\sim 14.2$ to $\sim 9.3$. The outburst, 
shown in Fig.\ref{aavso}, 
was fairly normal for a ``wide'' outburst, however, 
it followed
a quiescence period of only 69 days (the previous outburst was a 
short 7-day outburst that started  on September 14). The {\it HST}/COS 
observations of U Gem were taken on December 20 \& 26, 2012,
and on January 7 \& 30, 2013, following the wide November outburst.

\begin{deluxetable*}{cccccc}[t!] 
\tabletypesize{\scriptsize} 
\tablewidth{0pt}
\tablecaption{U Gem Cos G140L (1280\AA ) Spectra Observation Log}
\tablehead{
Epoch  & DATAID     & Obs. date  & Obs. time (UT) & Exp.time  & days$^1$    \\  
Number &            & yyyy-mm-dd & hh:mm:ss       & sec       &  
}
\startdata
1       & LC1U01010 & 2012-12-20 & 11:54:12        & 5761.6   &   15      \\
2       & LC1U02010 & 2012-12-26 & 04:59:28        &  960.0   &   21      \\
3       & LC1U06010 & 2013-01-07 & 23:26:05        &  960.1   &   33      \\
4       & LC1U04010 & 2013-01-30 & 22:03:00        &  960.1   &   56      \\
\enddata
\tablecomments{
Note: (1) In the sixth column we indicate the days counted since
the peak of the outburst that started on November 29, 2012 and lasted
15 days.   
} 
\end{deluxetable*}

The time at which the observations were taken is marked on 
Fig.\ref{aavso} 
with vertical arrows pointing down. The observation log is given 
in Table 2.

COS was set up in the FUV configuration with the G140L 
grating centered at 1280 \AA\ in time tag mode. The data were collected
through the primary science aperture (PSA) of 2.5 arcsec diameter. 
This configuration
produces two spectral segments covering $\sim$915-1193 \AA\ and 
$\sim$1266-1248 \AA , with a gap between 1193~\AA\ and 1266~\AA . 
While the resulting spectrum covers the Lyman 
series down to its cut-off, it does not cover the L$\alpha$ region.

\begin{figure}[h!] 
\vspace{-14.cm} 
\gridline{\fig{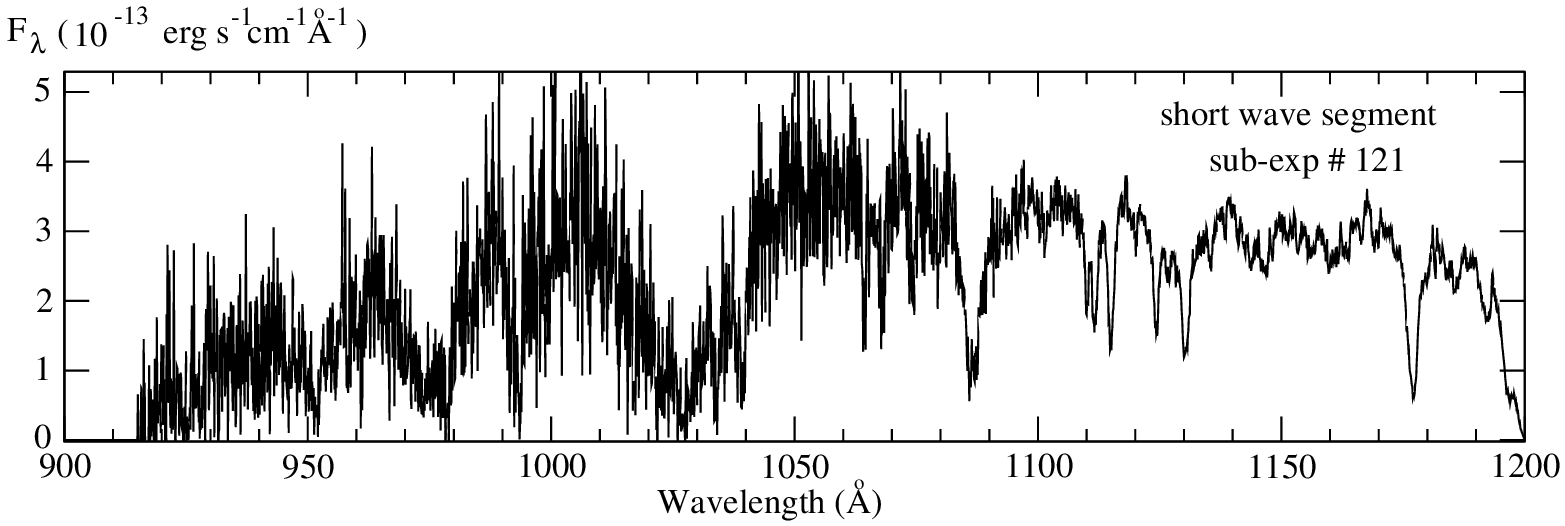}{0.6\textwidth}{}
          }
\vspace{-5.5cm} 
\gridline{\fig{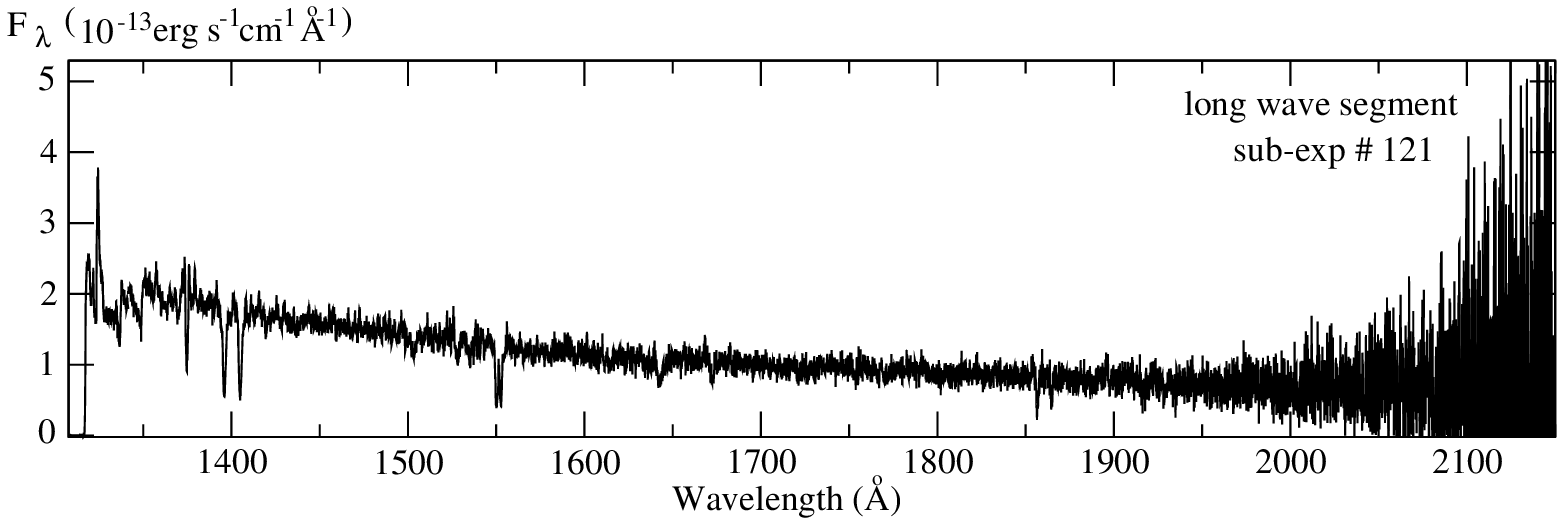}{0.6\textwidth}{}
          }
\vspace{-7.cm} 
\caption{
A sub-exposure (from a single detector position) is presented here 
to show the poor signal in the very short and very long wavelengths 
(sub-exposure \#1 of ({\it HST}) orbit \#2  obtained at epoch \#1). 
In this spectrum the gap occurs between 1195~\AA\ and 1318~\AA\ ,
though some detector edge effects affect the spectrum and makes the
actual gap wider. We discard the region $> 2000$~\AA , however,
it is likely that the spectrum
is affected at even shorter wavelengths ($\sim 1900$~\AA ). 
\label{rawsubex} 
}
\end{figure}

The data were processed with CALCOS version 3.1.8 through the pipeline
which produces for each {\it HST} orbit 4 sub-exposures generated
by shifting the position of the spectrum on the detector by 
20~\AA\ each time. This strategy is to reduce detector effects. 
It is left to the observer to compare the 4 sub-exposures and identify 
all the detector artifacts which appear on all sub-exposures but with 
a shift of 20~\AA\ from sub-exposure to sub-exposure.  
The first epoch data are covering 3 {\it HST} orbits and consist     
of 12 sub-exposures, the remaining data obtained at epochs 2, 3 and 4
all consist of one {\it HST} orbit each, made of 4 sub-exposures each. 
The second sub-exposure of epoch \#2 was lost due to the 
COS light path being blocked during that exposure. 
We therefore collected a total of 23 sub-exposures each consisting
of two spectral segments with a gap in the L$\alpha$ region. 
In the first sub-exposure of each exposure(/{\it HST} orbit), the 
gap spans the region  $\lambda \sim [ 1195-1318]$\AA ,
in the second sub-exposure $\lambda \sim [ 1175-1298]$\AA ,      
in the third sub-exposure $\lambda \sim [ 1155-1278]$\AA ,      
and in the fourth and last sub-exposure $\lambda \sim [ 1135-1258]$\AA .      
The extracted spectra have a signal that deteriorates toward the edges of the 
spectrum (see Fig.\ref{rawsubex}).

\begin{figure*}
\vspace{-7.cm} 
\gridline{\fig{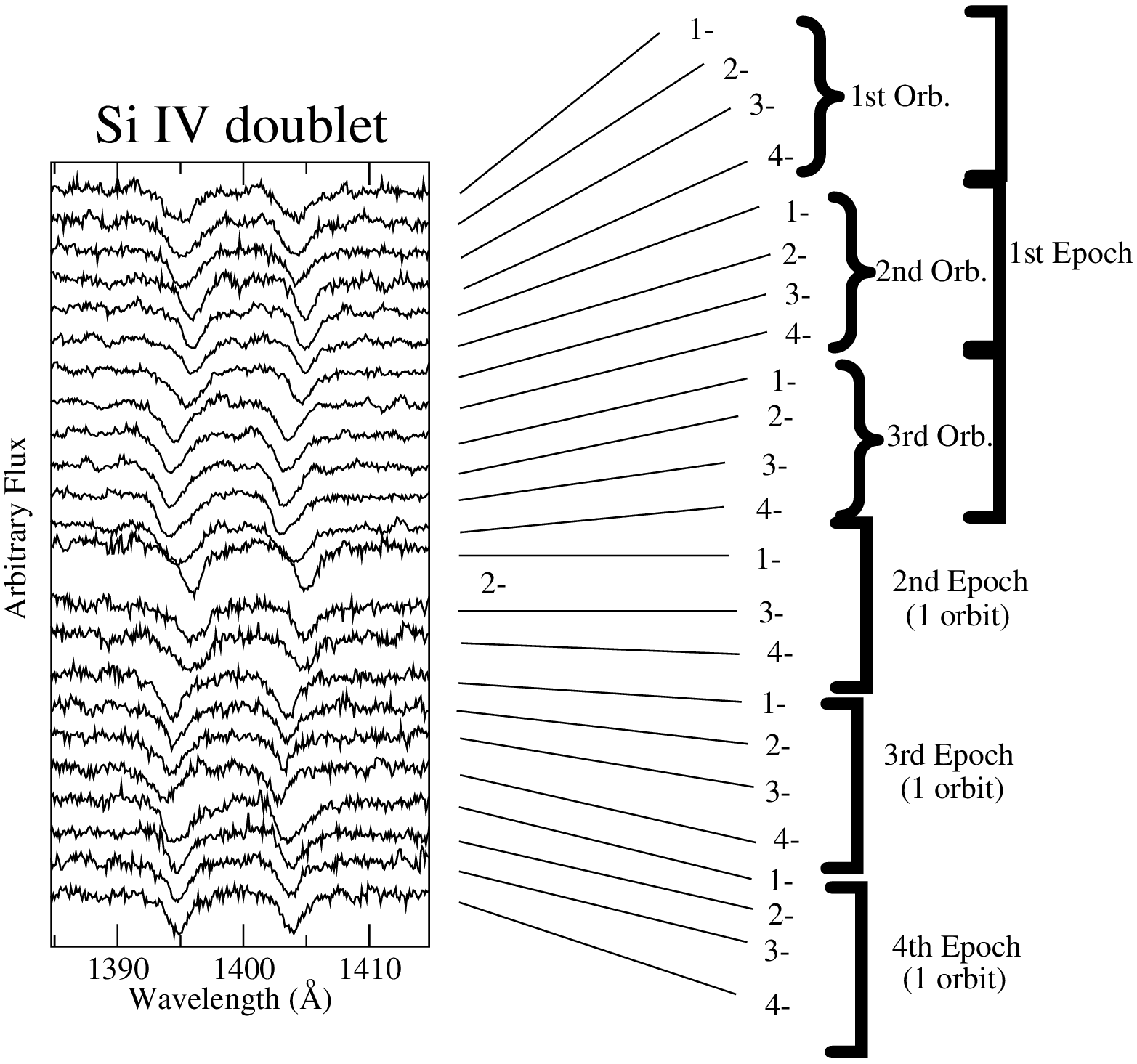}{0.8\textwidth}{} }
\vspace{-1.5cm} 
\caption{
The Si\,{\sc iv} doublet (1392.8, 1402.8) is shown 
for the individual sub-exposures spectra. 
The {\it HST}/COS data were obtained at 4 epochs, covering 
6 {\it HST} orbits as shown on the right. 
Each {\it HST} orbit generated 4 sub-exposures (one for each of 
the 4 COS detector position) totaling 23 spectra      
(sub-exposure 2 of the second epoch contains no data).  
The sub-exposures of the 2nd and 3rd orbits obtained at
the first epoch have an exposure time of $\sim$600 s each, 
the remaining sub-exposures have an exposure of only 
$\sim$200 s each. The difference in the S/N can be seen in   
the figure.
\label{si4doublet}  
}
\vspace{-9.cm}
\plotone{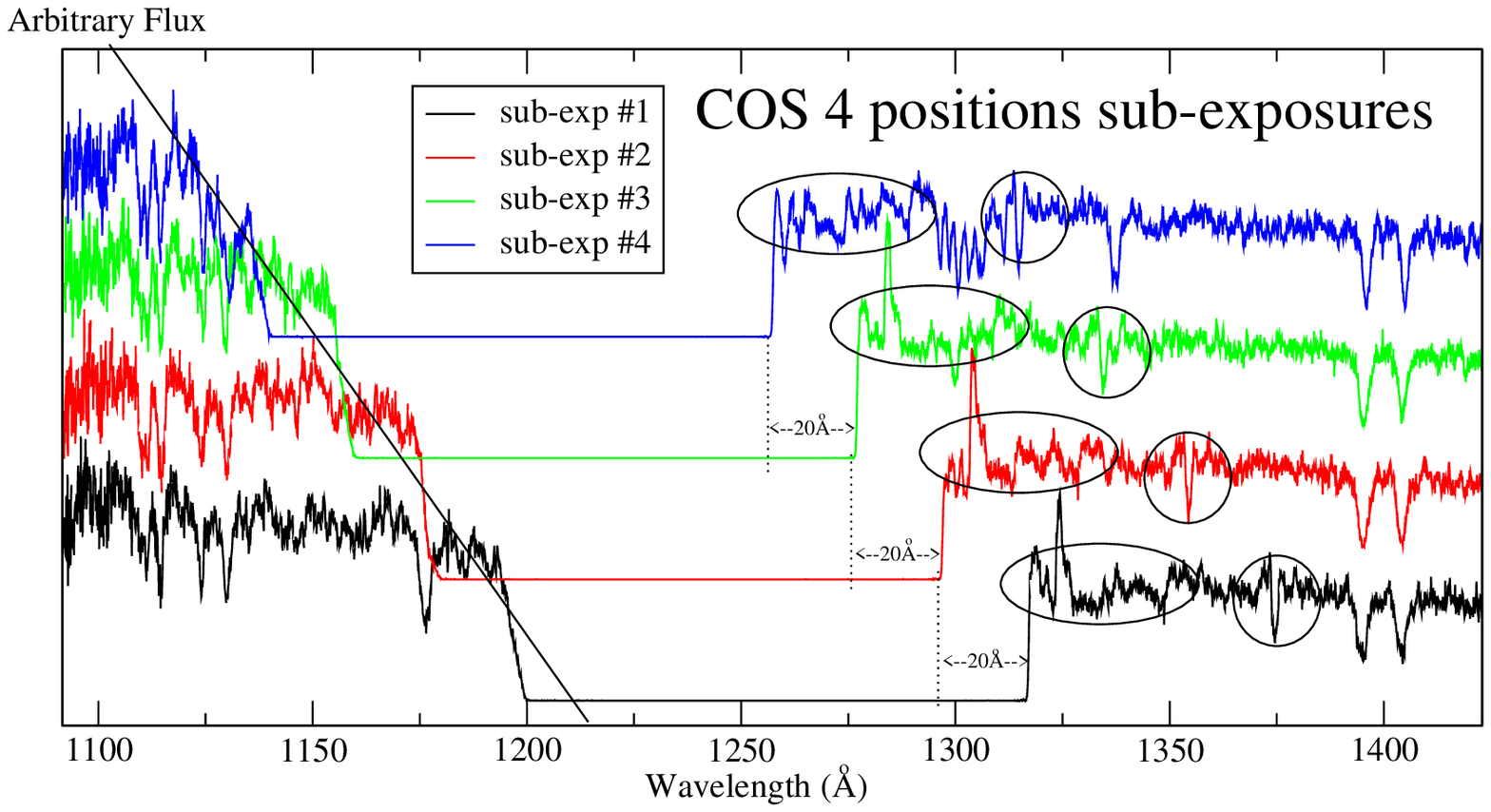} 
\vspace{-3.5cm}
\caption{
Each COS exposure is made of 4 sub-exposures that are generated by shifting
the position of the spectrum on the detector by 20~\AA . This strategy allows to
identify and compensate for detector artifacts by inspecting the 4 sub-exposures
simultaneously. We display here the first {\it HST}/COS observations sub-exposures
vertically shifted for clarity. Detector artifacts repeat themselves at 20~\AA\ 
intervals and are easily found by looking for similar pattern shifted by 20~\AA\
from sub-exposure to sub-exposure. In the present case, for illustration,
we isolate two such 
artifacts within the circle and within the ellipse. Unfortunately the detector
artifacts are stronger toward the edges of the detectors which render the 
effective detector gap   larger than it actually is.  
\label{detectors} 
}
\end{figure*}

For the first orbit of epoch \#1, and for the single orbit of 
epoch \# 2, 3, and 4, the effective good exposure time 
of each sub-exposure is a little more than $\sim$200 sec.
For the second and third orbits of epoch \#1 the exposure time
for each sub-exposure is $\sim 600$ sec.    
In Fig.\ref{si4doublet} we display 
the 23 spectra extracted from the sub-exposures
in the Si\,{\sc iv} doublet ($\sim 1400$~\AA ) region to show 
the quality of the spectra.  

Due to the gap between the two segments of the detector and the 20~\AA\ shift 
in the detector position for each of the 4 sub-exposures, and due to detector
artifacts (especially near its edges), the spectra are not reliable 
beyond the C\,{\sc iii} (1175) absorption lines, nor below $\sim$1300~\AA .
In Fig.\ref{detectors} 
we illustrate how the detector edges (and gain sag) affect  
some of the spectral regions, and we explain how we identify these artifacts 
that need to be discarded. 
For this reason the spectral lines analysis is carried out on the individual
sub-exposures, which have a much lower S/N than the combined spectrum.

Because of the sub-exposure spectra having a different wavelength
coverage, the C\,{\sc iii} (1175) and N\,{\sc iii} (1183) absorption
lines are only present in 6 sub-exposures.

\subsection{{\bf The {\it FUV} Absorption Lines} }

\begin{figure}[h] 
\vspace{-2.cm} 
\plotone{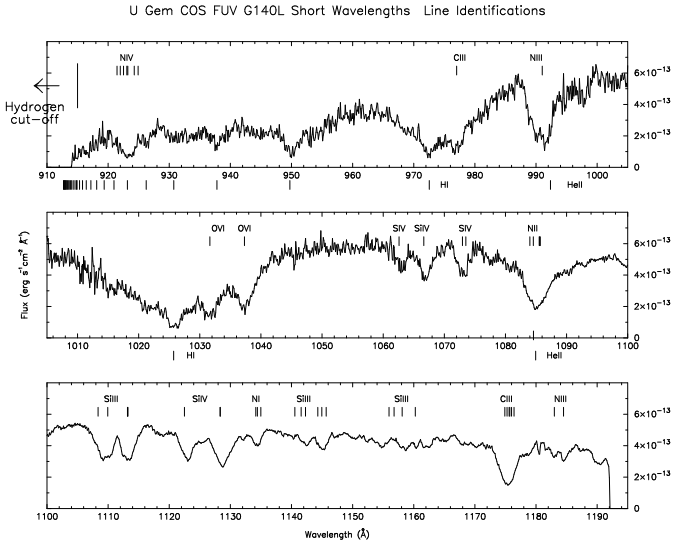}          
\caption{The short wavelength region of the COS G140L spectrum 
of U Gem, exposure LC1U01010 (epoch \# 1),  is shown  here for line identifications.  
The Hydrogen Lyman series lines are marked underneath the panels together
with the 
He\,{\sc ii} Balmer lines $n=5$ at 1084.94~\AA\ and $n=7$ at 992.36~\AA .
The spectrum is characterized by the broad wings of the hydrogen Lyman 
series in absorption as well as broad absorption lines from low 
and high ionization species. The carbon, silicon, sulphur, and nitrogen
lines are typical to accreting WD          stars, though the nitrogen 
lines seems to be stronger than usual. The N\,{\sc iv} and 
O\,{\sc vi} doublet possibly form
in a hot ($\sim 80,000$~K) 
medium veiling the WD. 
The spectrum does not show any ISM absorption lines and clearly reveals  
a moderately hot accreting WD.     
\label{short} 
} 
\end{figure}

\begin{figure}[h] 
\vspace{-2.cm} 
\plotone{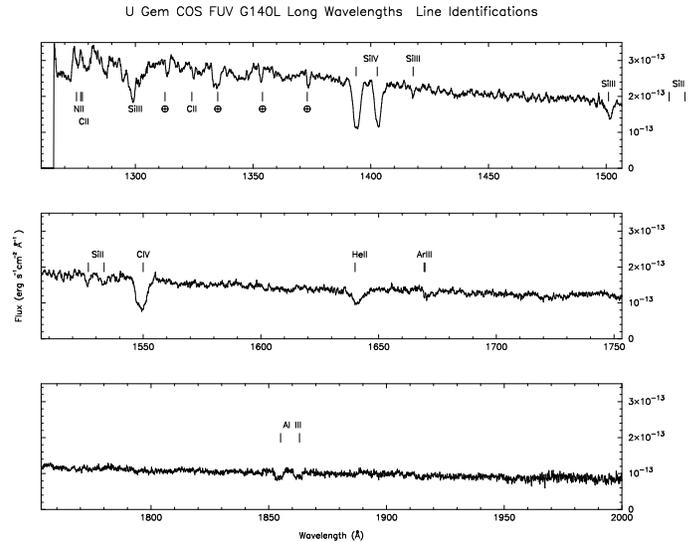}           
\caption{The COS spectrum (as in Fig.\ref{short}, 
exp.LC1U01010) is shown in the  
long wavelength region, but on a different vertical scale. 
The spectrum reveals the usual absorption
lines seen the photospheres of accreting WDs, such as the 
Si\,{\sc iii} (1300), C\,{\sc ii} (1335), silicon doublet ($\sim$1400),
C\,{\sc iv} (1550) and He\,{\sc ii} (1638). 
The spectrum exhibits some unusually strong Ar\,{\sc iii}
($\sim$1669) and Al\,{\sc iii} lines
(1353, 1855, \& 1863). 
The absorption features marked with the $\oplus$ symbol  are detector
artifacts (gain sag in regions illuminated by airglow). 
In the longer wavelengths (lower panel),
the spectrum becomes extremely noisy and for that reason we discard
the region $> 2000$~\AA .  
\label{long} 
} 
\end{figure}

\begin{deluxetable}{ll}[h!] 
\tabletypesize{\scriptsize} 
\tablewidth{0pt}
\tablecaption{Absorption Lines in the COS Spectra of U Gem in quiescence} 
\tablehead{
Ion                     & $\lambda$(\AA)    
}
\startdata
 N\,{\sc iv}                & 921.46-924.91; 923.2   \\  
 C\,{\sc iii}               & 977.0   \\  
 N\,{\sc iii}               & 989.8, 991.56   \\  
 He\,{\sc ii}               & 992.4   \\  
 O\,{\sc vi}                & 1031.9    \\  
 O\,{\sc vi}                & 1037.6   \\  
 S\,{\sc iv}                & 1062.7   \\ 
 Si\,{\sc iv}               & 1066.6   \\ 
 S\,{\sc iv}                & 1073.0   \\  
 He\,{\sc ii}               & 1084.9   \\
 N\,{\sc iii}               & 1085.5   \\  
 Si\,{\sc iii}              & 1108.4, 1109.9, 1113.2     \\ 
 Si\,{\sc iv}               & 1122.5, 1128.3    \\ 
 Si\,{\sc iii}              & 1140.6-1145.7; 1144.3    \\ 
 C\,{\sc iii}               & 1174.6-1176.8; 1175.3 \\  
 N\,{\sc iii}$^*$           & 1183.0   \\  
 N\,{\sc iii}$^*$           & 1184.5-1185.6   \\  
 Si\,{\sc iii}              & 1294.4, 1296.7, 1298.9, 1301, 1303.2  \\ 
 C\,{\sc ii}                & 1323.9-1324.0 \\  
 C\,{\sc ii}$^*$            & 1334.5, 1335.7  \\ 
 Si\,{\sc iii}              & 1341.5, 1342.4, 1343.4  \\   
 Si\,{\sc iv}               & 1392.8, 1402.8 \\  
 Si\,{\sc iii}              & 1417.2 \\ 
 Si\,{\sc iii}              & 1500.2, 1501.2, 1501.9 \\  
 Si\,{\sc ii}               & 1526.7, 1533.4  \\  
 C\,{\sc iv}                & 1548.2, 1550.8  \\  
 He\,{\sc ii}               & 1640.3, 1640.5 \\ 
 Ar\,{\sc iii}              & 1669.3, 1669.7    \\ 
 Al\,{\sc iii}              & 1854.7, 1862.8  \\ 
 Fe                         & $\sim$1450 - 1600  \\ 
\enddata
\tablecomments{All the lines listed here have been marked in 
Figs.\ref{short} and \ref{long}. 
As described in the analysis section, many lines were identified
by comparing them to low-metallicity synthetic stellar spectra 
while increasing the abundance of one species (metal) at a time. 
In this manner we identified the sawtooth pattern between 
$\sim 1450$~\AA\ and $\sim 1600$~\AA\ as a multitude of small iron
absorption lines (Fe\,{\sc ii}).
Lines anotated with an asterisk (*) are contaminated with detector 
artifacts.}  
\end{deluxetable}

In this subsection we present the absorption lines
detected  in the COS spectrum of U Gem. While many lines can be 
easily identified, some other lines were identified during our spectral
analysis and the procedure we follow for line identifications is 
described in the analysis section. We did not detect any emission
features in the COS spectrum of U Gem.    

During quiescence we expect the WD to dominate the FUV 
and, therefore, we expect the COS spectrum to exhibit       
absorption lines from a moderately hot ($\sim$30,0000-40,000 K) WD,
namely the hydrogen Lyman lines, helium Balmer lines, 
low ionization species lines, as well as possibly some
higher ionization species lines.  
The individual sub-exposure spectra are very noisy, especially 
below 1000~\AA, and we consequently 
first proceed by combining together the sub-exposures of  
the first epoch to clearly display the absorption lines and identify them 
below 1080~\AA.
For clarity the resulting epoch \#1 spectrum is presented in two figures.  
The actual spectral analysis is carried out on the sub-exposures
rather than on the combined spectrum. 

In Fig.\ref{short} we display the short wavelength segment
which has a similar spectral range to 
{\it FUSE} spectra. 
The hydrogen 
Ly$\beta$ absorption line is the main broad feature dominating the spectrum 
in the middle panel. The hydrogen 
Ly$\gamma$ and Ly$\delta$ absorption features are also
clearly apparent, but the higher orders of the series are not. 
The spectrum is marked by broad absorption features usually detected in the
{\it FUSE} spectra of CV WDs \citep{god12} from 
carbon (C\,{\sc iii} 977, 1175), 
nitrogen (N\,{\sc iii} 990, 1085, 1184), 
silicon (Si\,{\sc iii} 1110, 1144; Si\,{\sc iv} 1167, 1122, 1128),
and sulphur (S\,{\sc iv} 1063, 1073).    
In addition, some higher ionization species absorption lines appearing
only at higher temperature ($T\sim80,000$~K) are also seen, 
N\,{\sc iv} (923) and the O\,{\sc vi} doublet (1032, 1038), 
an indication that a hot absorbing medium is in front of the WD. 
The S\,{\sc vi} (933, 944) and P\,{\sc v} (1118, 1128) absorption lines, 
previously seen in the {\it FUSE}
spectra of U Gem in quiescence \citep{lon06}, 
are not detected here in the spectrum. 
The line identifications process is explained
in the analysis section where each spectrum (and its sub-exposures) 
are modeled with realistic photospheric WD models.  
We list the absorption lines in Table 3 where only the rest (laboratory)
wavelength is listed for each line, since the exact position of the lines 
changes for each sub-exposure.   

Not unsurprizingly the COS spectrum of U Gem does not show any 
interstellar absorption lines as it is a nearby system.  

The long wavelength segment of the epoch \# 1 spectrum is displayed in
Fig.\ref{long}, where, for clarity, the vertical scale has been stretched.  
The first striking feature of the long wavelength segment spectrum are
the numerous absorption features seen at wavelengths below 1400~\AA . 
An inspection of the sub-exposures reveals that most of the `lines' 
are due to detector artifacts (gain sag in regions illuminated by airglow), 
as they are seen in each sub-exposure shifted by 20~\AA . 
Below 1400~\AA, we detect with confidence Si\,{\sc iii} ($\sim$1300)  
and C\,{\sc ii} (1324) lines. The C\,{\sc ii} (1335) is contaminated with 
a detector artifact. All the other features between 1265~\AA\ and 1380~\AA\  
are detector artifacts. Fortunately, the longer wavelength region is not
affected and we identify silicon lines (Si\,{\sc iv} doublet $\sim$1400; 
Si\,{\sc iii} $\sim$1342, 1417, 1500; Si\,{\sc ii} 1527, 1533), the 
C\,{\sc iv} (1550) line, the He\,{\sc ii} (1640) line, as well as 
aluminum (Al\,{\sc iii} 1855, 1863) and possibly argon (Ar\,{\sc iii} 1670) 
lines. 
In the line analysis section we will show that the sawtooth pattern observed 
from $\sim 1420$~\AA\ to $\sim$1630~\AA\ is likely due to iron (Fe\,{\sc ii}).

\section{{\bf Spectral Analysis Results and Discussion} }

We use Ivan Hubeny's FORTRAN suite of codes TLUSTY and SYNSPEC
\citep{hub88,hub95} to generate synthetic spectra  
for high-gravity stellar atmosphere models.   
We first generate a one-dimensional vertical
stellar atmosphere structure with TLUSTY for a given surface gravity $\log(g)$,
effective surface temperature $T_{\rm eff}$, and surface composition 
of the star. We treat hydrogen and helium explicitly, 
and treat nitrogen, carbon and oxygen implicitly 
\citep{hub95}. 
We then run SYNSPEC using the output stellar atmosphere model from TLUSTY
as an input, and generate a synthetic stellar spectrum over a given
(input) wavelength range, here  from 900~\AA\  to 3200~\AA . 
The code SYNSPEC derives the detailed radiation and flux distribution of 
continuum and lines and generates the output spectrum.
For temperatures above 35,000~K the approximate NLTE treament of lines 
is turned on in SYNSPEC \citep{hub95}.           
Rotational and instrumental broadening as well as limb darkening 
are reproduced using the routine ROTIN. 
For full details of the codes see \citet{hub17a,hub17b,hub17c}.   
 
Following this procedure, we generated photospheric models 
for U Gem in quiescence with an effective WD  
temperature ranging from 25,000~K to 50,000~K in increments of 
1,000~K, 500~K, 
and $\sim$250~K, as needed for the convergence of the model fit.  
The effective surface gravity was set to $log(g)=8.8$ 
to match the large mass of the WD $\sim 1.1-1.2M_{\odot}$. 
We varied the projected stellar rotational velocity $V_{rot} \sin(i)$  
from 100~km~$s^{-1}$ to 250~km~$s^{-1}$ to match the known WD velocity
of about $100-150$~km~s$^{-1}$, 
and in order to fit the absorption features of the spectrum, we varied  
the abundances of the elements, one at a time,  
from 0.01$\times$ solar to 20$\times$ solar. 

The spectral analysis to derive the effective WD temperature and the 
abundance of the elements was  carried out iteratively. 
Namely, we first assessed the WD temperature
using solar abundances, then for that temperature we varied
the abundances one element at a time, then we fine-tuned the WD
temperature using the best-fit abundances.   

\subsection{{\bf Abundances}}

\begin{table*}[t!] 
\caption{Abundance of Elements in U Gem COS Spectra in quiescence} 
\begin{center} 
\begin{small} 
\begin{tabular}{rcccccccccccccc}
\hline      
      Ion      & C\,{\sc iii} & C\,{\sc iv} & N\,{\sc iii} & N\,{\sc iii} & N\,{\sc iii} & S\,{\sc iv} & Si\,{\sc iv} & Si\,{\sc iii} & Si\,{\sc iv} & Si\,{\sc iii} & Si\,{\sc iii} & Si\,{\sc iv}  & Al\,{\sc iii} &  Orb.Phase \\
$\lambda$(\AA )& 1175*        & 1550        & 990          & 1085*        & 1184*        & 1063-73     & 1067         & $\sim$1112   & $\sim$1126    & $\sim$1143    & $\sim$1160*   & $\sim$1400    & 1856-64       &  $\phi$    \\ 
\hline      
Exp.           &              &             &              &              &              &             &              &              &                &              &               &              &                &            \\     
\hline      
1-1-1          & 0.4          & 0.75        & 20           & 0.01         & 0.4          & 0.5         & 0.1          & 1.0          & 0.2            & 0.3          & 0.7           & 0.8          & 1.0           & 0.51      \\  
    2          & -            & 2.0         & 20           & 5            & -            & 0.2         & 0.5          & 2.0          & 0.2            & 0.3          & 0.7           & 1.5          & 1.0           & 0.54      \\  
    3          & -            & 0.75        & 20           & 10           & -            & 0.5         & 0.4          & 1.0          & 0.1            & 0.3          &  -            & 1.0          & -             & 0.57       \\  
    4          & -            & 0.70        & 20           & 20           & -            & 3.0         & 2.0          & 3.0          & 1.0            &  -           &  -            & 1.0          & 20            & 0.78       \\  
  2-1          & 0.4          & 1.0         & 20           & 20           & 0.1          & 1.0         & 0.1          & 5.0          & 0.6            & 0.4          & 0.5           & 1.0          & 20            & 0.81      \\  
    2          & -            & 0.9         & 20           & 1            & -            & 0.8         & 0.1          & 3.0          & 0.3            & 0.3          & 0.5           & 0.5          & 0.8           & 0.86     \\  
    3          & -            & 0.5         & 20           & 20           & -            & 0.6         & 0.05         & 2.0          & 0.1            & 0.3          &   -           & 0.9          & 0.5           & 0.91    \\  
    4          & -            & 1.0         & 20           & 10           & -            & 0.4         & 0.5          & 2.0          & 0.4            &  -           &   -           & 1.0          & 10            & 0.16  \\  
  3-1          & 0.6          & 1.0         & 20           & 20           & 0.05         & 1.0         & 0.1          & -            & 1.0            & 0.3          & 0.5           & 3.0          & 10            & 0.21   \\  
    2          & -            & 1.0         & 20           & 20           & -            & 1.5         & 0.05         & -            & 1.0            & 0.3          & 0.5           & 1.0          & 20            & 0.25     \\  
    3          & -            & 1.0         & 20           & 20           & -            & 0.8         & 0.05         &              &                &              &   -           & 2.0          & 20            & 0.30   \\  
    4          & -            & 2.0         & 20           & 10           & -            & 0.5         & 0.05         & 1.0          & 0.2            &  -           &   -           & 2.0          & 1.0           & 0.53    \\  
2-1-1          & 0.2          & 1.5         & $>>1$        & 5            & 0.7          & 1           & 0.4          & 3.0$^*$      & 0.4$^*$        & 0.3$^*$      & 0.5$^*$       & 1.0          & 7             & 0.80  \\  
    2          & -            & -           & -            & -            & -            & -           & -            & -            & -              &  -           &   -           & -            & -             & 0.83   \\  
    3          & -            & 0.05        & 10           & 0.5          & -            & -           & -            & 0.6$^*$      & 0.08$^*$       & 0.2$^*$      &   -           & 0.7          & 1             & 0.85   \\  
    4          & -            & 0.3         & 1.0          & 0.5          & -            & 1           & 0.2          & 0.6          & 0.15           &  -           &   -           & 0.5          & -             & 0.88    \\  
3-1-1          & 0.1          & 1.0         & 1.0          & 10           & 1.0          & 1           & 0.3          & 0.9          & 0.40           & 0.2          &   -           & 1.0          & 3             & 0.98   \\  
    2          & -            & 1.0         & -            & 3            & -            & 2           & 0.2          & 1.0          & 0.2            & 0.2          & 1.0           & 0.6          & 2             & 0.00   \\  
    3          & -            & 0.7         & -            & 1.0          & -            & 0.4         & -            & 0.7$^*$      & 0.1            &  0.2         &  -            & 0.5          & 0.2           & 0.03  \\  
    4          & -            & 0.1         & -            & 0.1          & -            & 0.4         & 0.5          & 0.4$^*$      & 0.15$^*$       &  -           &  -            & 0.6          & 1             & 0.24   \\  
4-1-1          & 0.5          & 2.5         & 20           & 20           & 0.5          & 1           & 0.2          & 1$^*$        & 0.2            & 0.2          &  -            & 2.0          & 3             & 0.66  \\  
    2          & -            & 2.5         & -            & 10           & -            & 1           & 0.2          & 1$^*$        & 0.15           & 0.15         & 0.4           & 1.0          & 1             & 0.69  \\  
    3          & -            & 1.5         & 10           & 10           & -            & 0           & -            & 1            & 0.08           & 0.3          &  -            & 0.7          & 1             & 0.96   \\  
    4          & -            & 3.0         & 10           & 10           & -            & -           & -            & 1            & 0.2            &  -           &  -            & 1.0          & 3             & 0.99    \\  
\hline
\end{tabular} 
\tablecomments{The lines were fitted with a rotational broadening
velocity of 150~km~s$^{-1}$. 
The values listed are accurate to the next adjacent discrete values 
(see text). 
The abundances derived from the 
C\,{\sc iii} 1175, N\,{\sc iii} 1185, N\,{\sc iii} 1184, and Si\,{\sc iii} 1160 line complexes
were deemed unreliable and are marked with an   
asterix * (top of the table). 
Unreliable abudances within the table are also marked with an asterix * .
No value is given for lines in the detector gap or in very noisy 
regions of the sub-exposure. 
Sub-exposure 2 of epoch 2 has no signal at all. 
}  
\end{small}  
\end{center}
\end{table*}

Except for hydrogen and helium that were set to solar abundances, 
we started by setting all the metals (i.e. $Z>2$) 
to a very low abundance such that 
no absorption lines formed in the theoretical spectra, 
except for the lines of the hydrogen Lyman and helium Balmer series.   
We then increased the abundance of one metal at a time starting 
at 0.01$\times$ solar, up to 20$\times$ solar, taking all the 
discrete values : 0.01, 0.02, 0.03, 0.04, 0.05, 0.08, 0.10, 0.12,
0.15, 0.2, 0.3, 0.4, 0.5, 0.6, 0.7, 0.8, 0.9. 1, 2, 3, 4, 5, 7, 10,
15, 20.    
In this manner we generated theoretical spectra with varied abundances  
for C, N, O, P, S, Si, Al, Ar, Fe, Mg, Na, and Ne, one at a time.
In all the models we choose $\log(g)=8.8$, and for the first epoch 
spectra the WD temperature was set (see next sub-section) to 41,500~K,
for the second epoch $T=39,850$~K, for the third epoch $T=37,650$~K,
and for the fourth and last epoch $T=36,350$~K.  
We then fitted each line in each single sub-exposure with the varied
abundances WD spectra until a match was found.

\begin{figure*}
\vspace{-4.1cm} 
\gridline{\fig{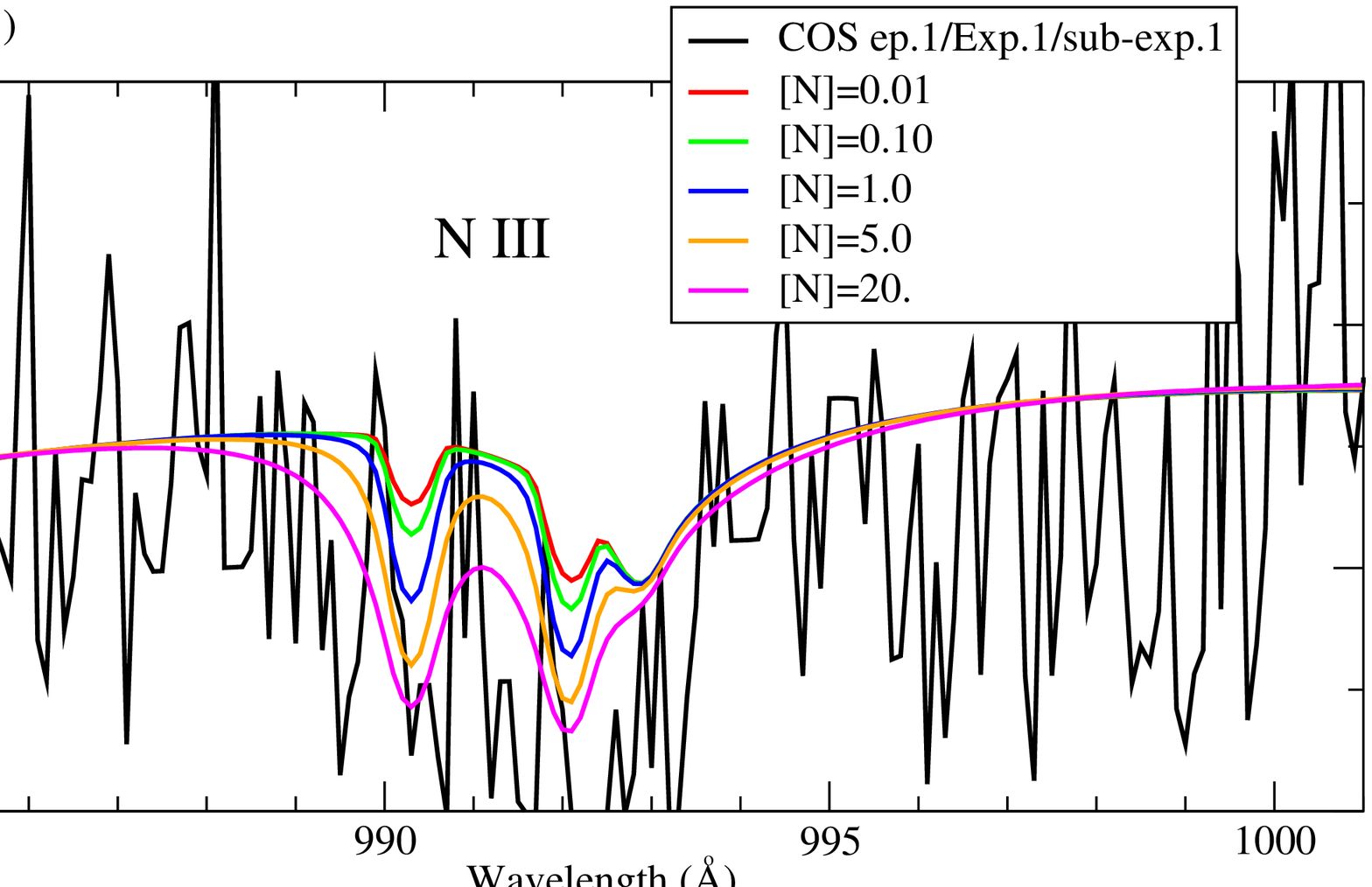}{0.33\textwidth}{(a)}
          \fig{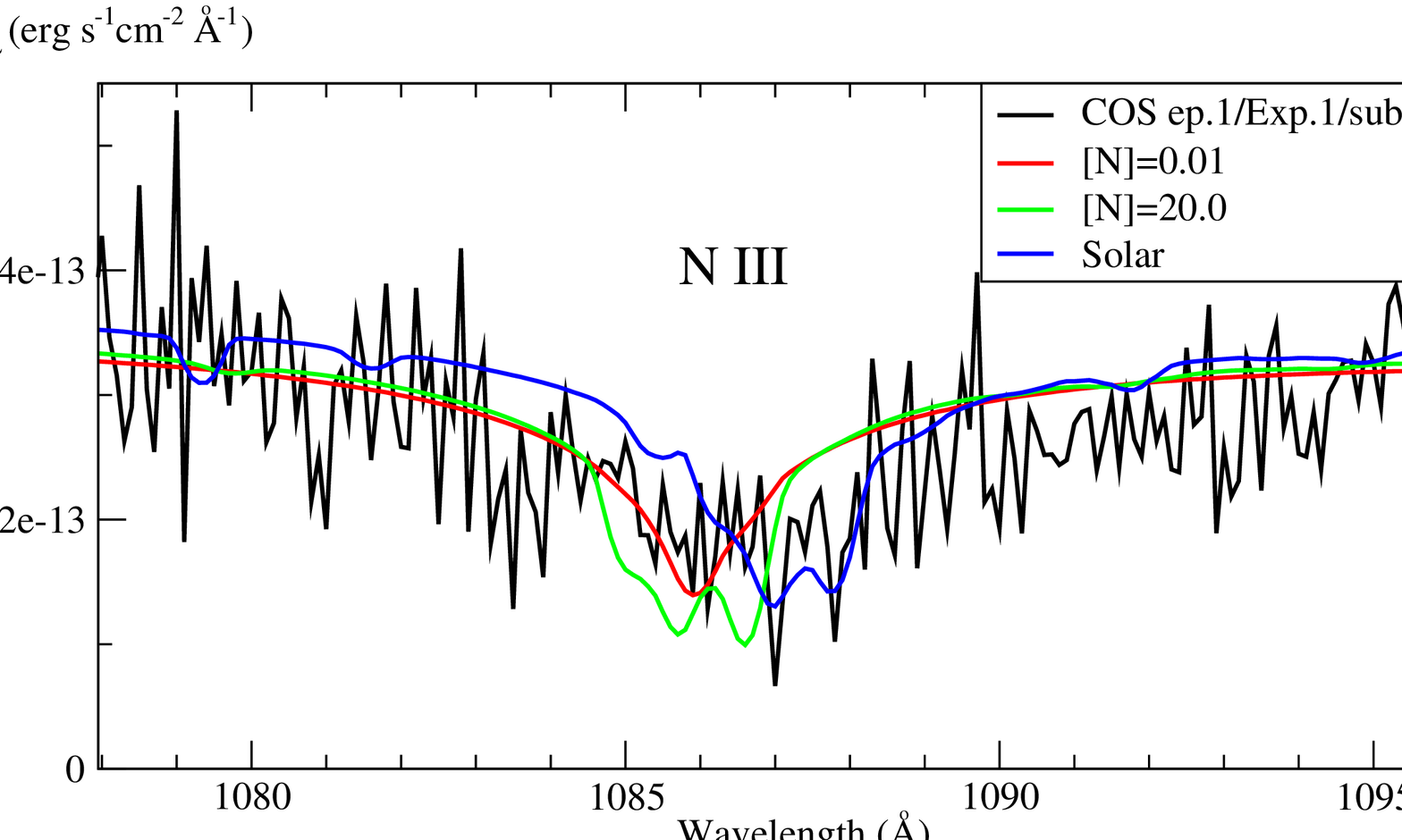}{0.33\textwidth}{(b)}
          \fig{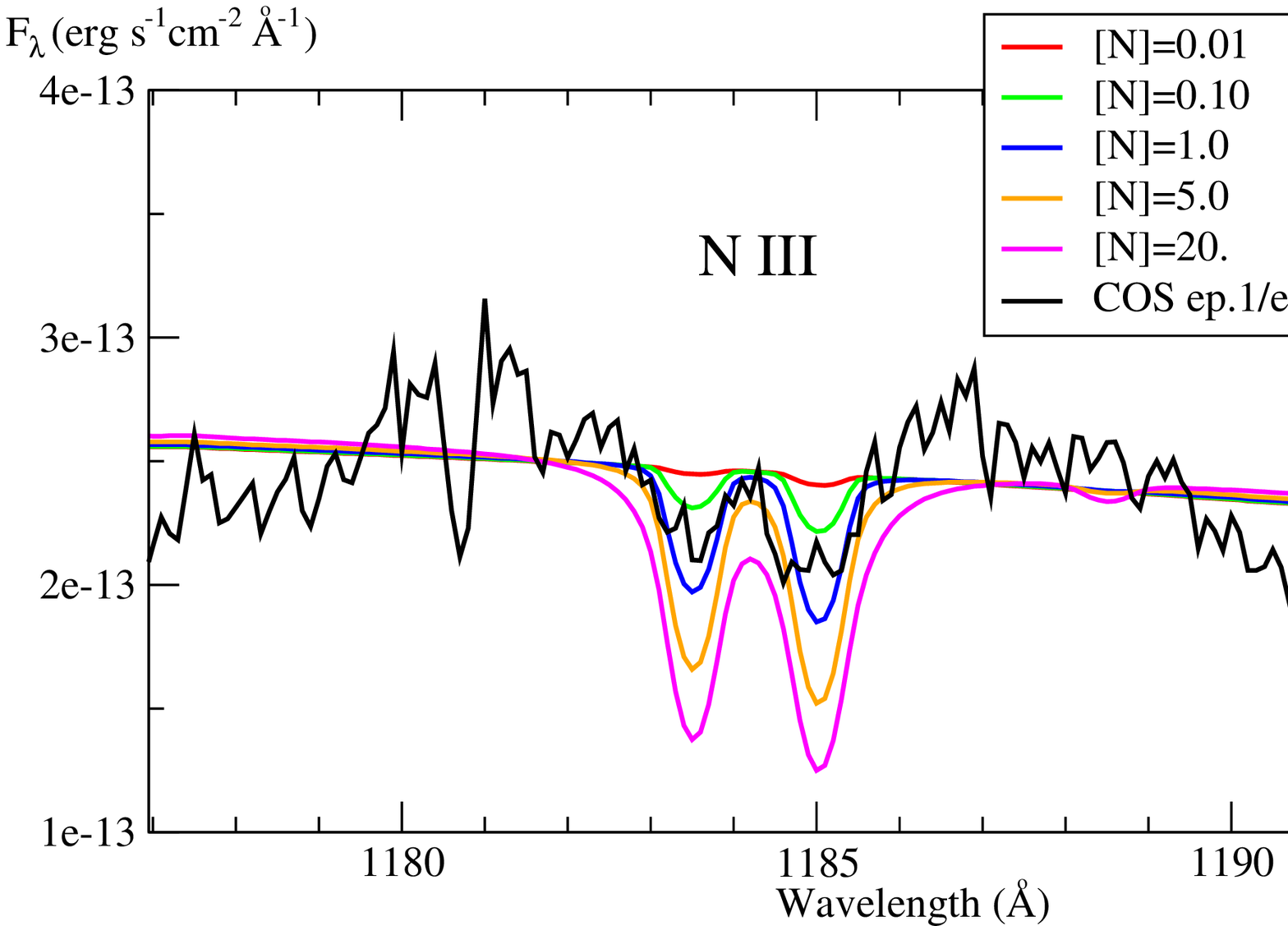}{0.33\textwidth}{(c)}
          }
\vspace{-2.7cm} 
\gridline{\fig{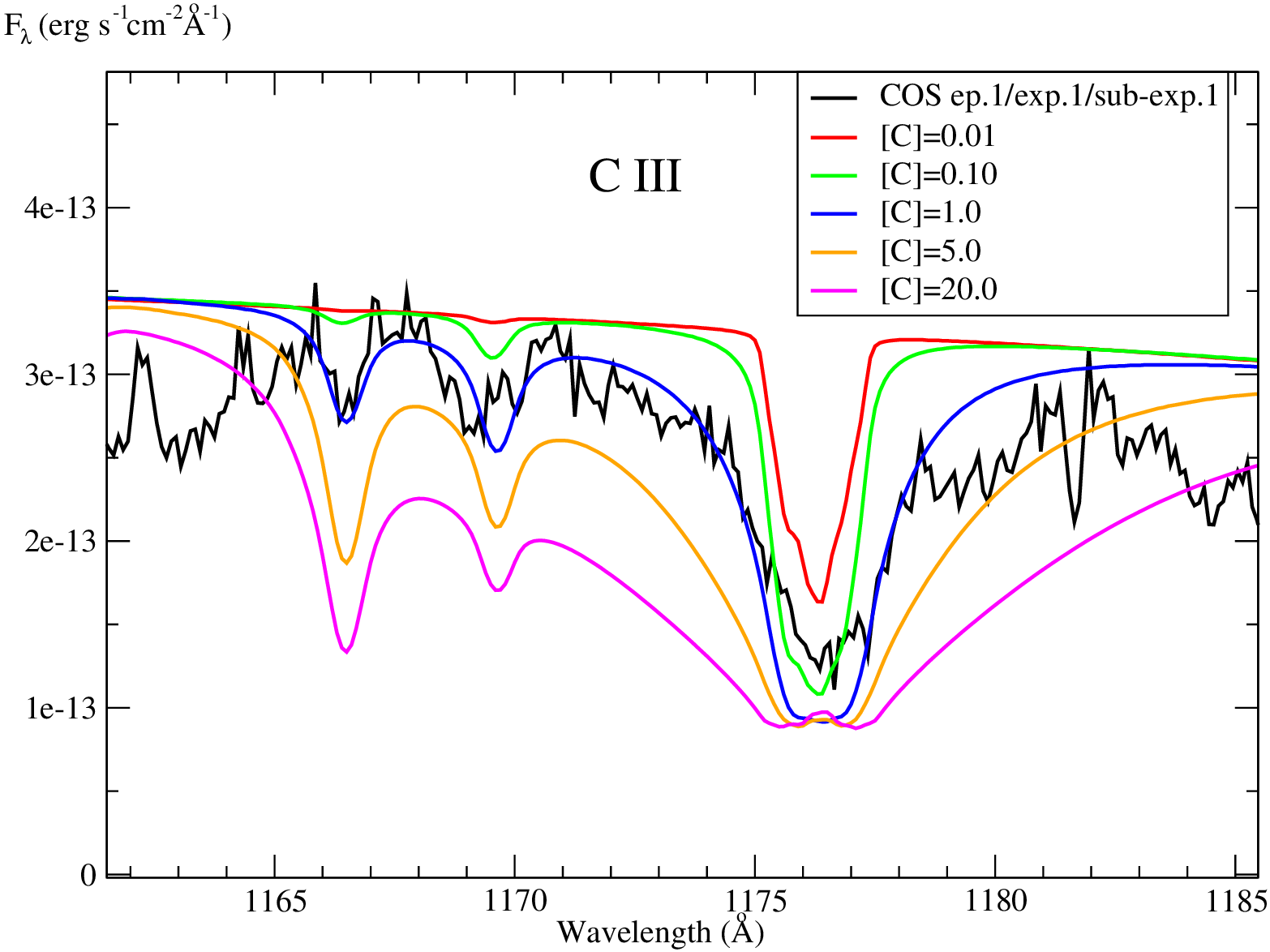}{0.30\textwidth}{(d)}
          \fig{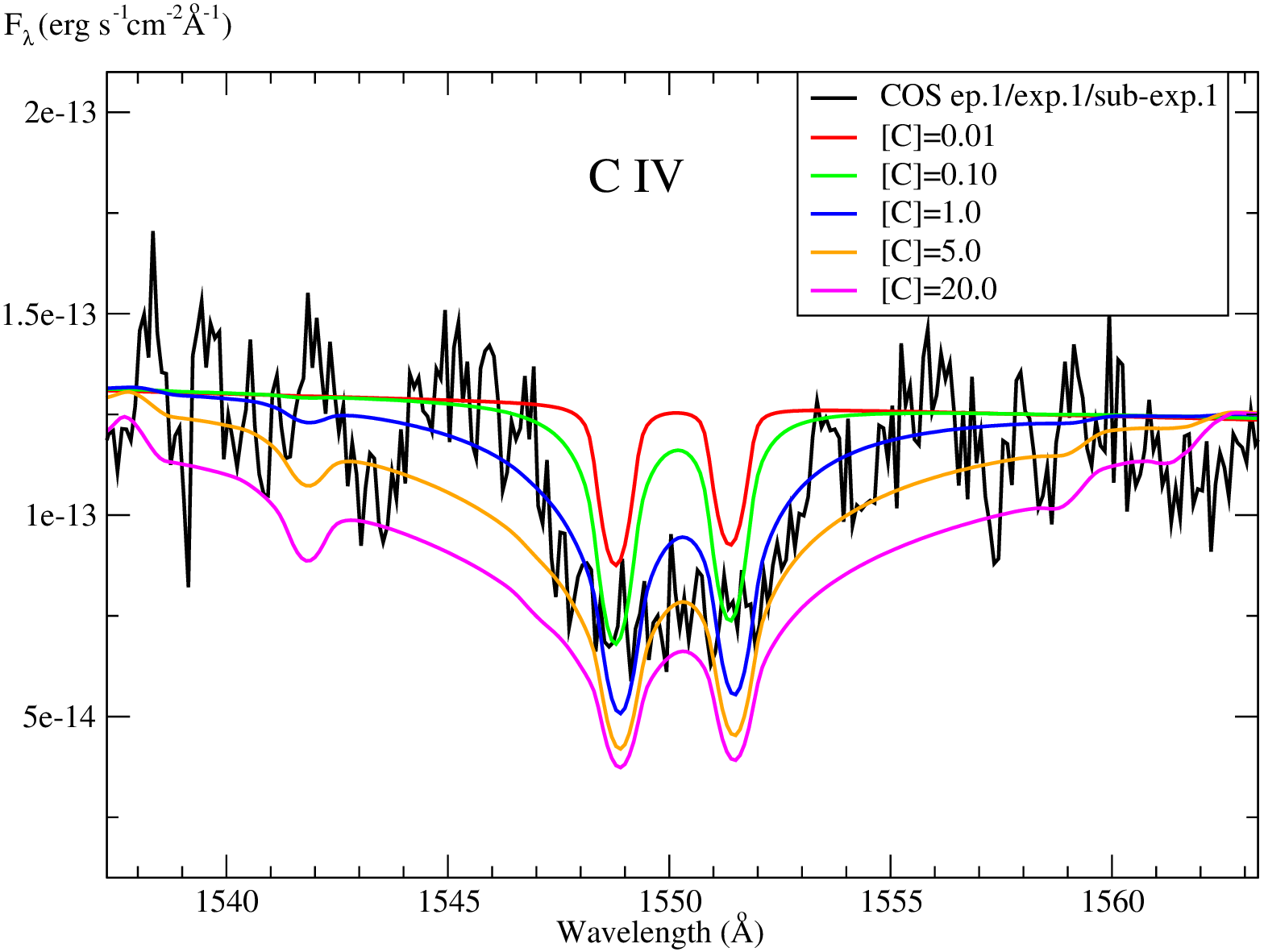}{0.30\textwidth}{(e)}
          \fig{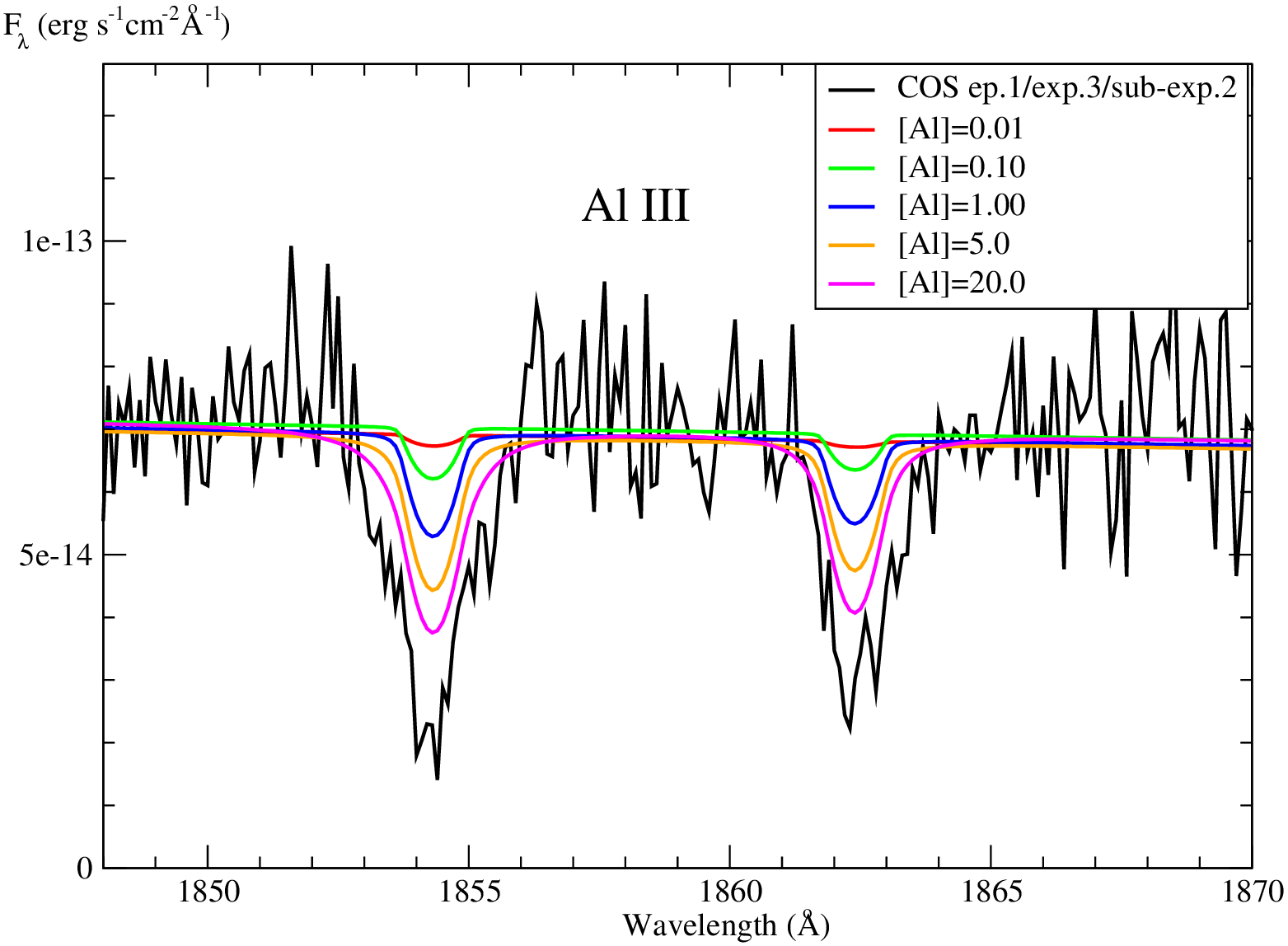}{0.30\textwidth}{(f)}
          }
\vspace{-2.7cm} 
\gridline{\fig{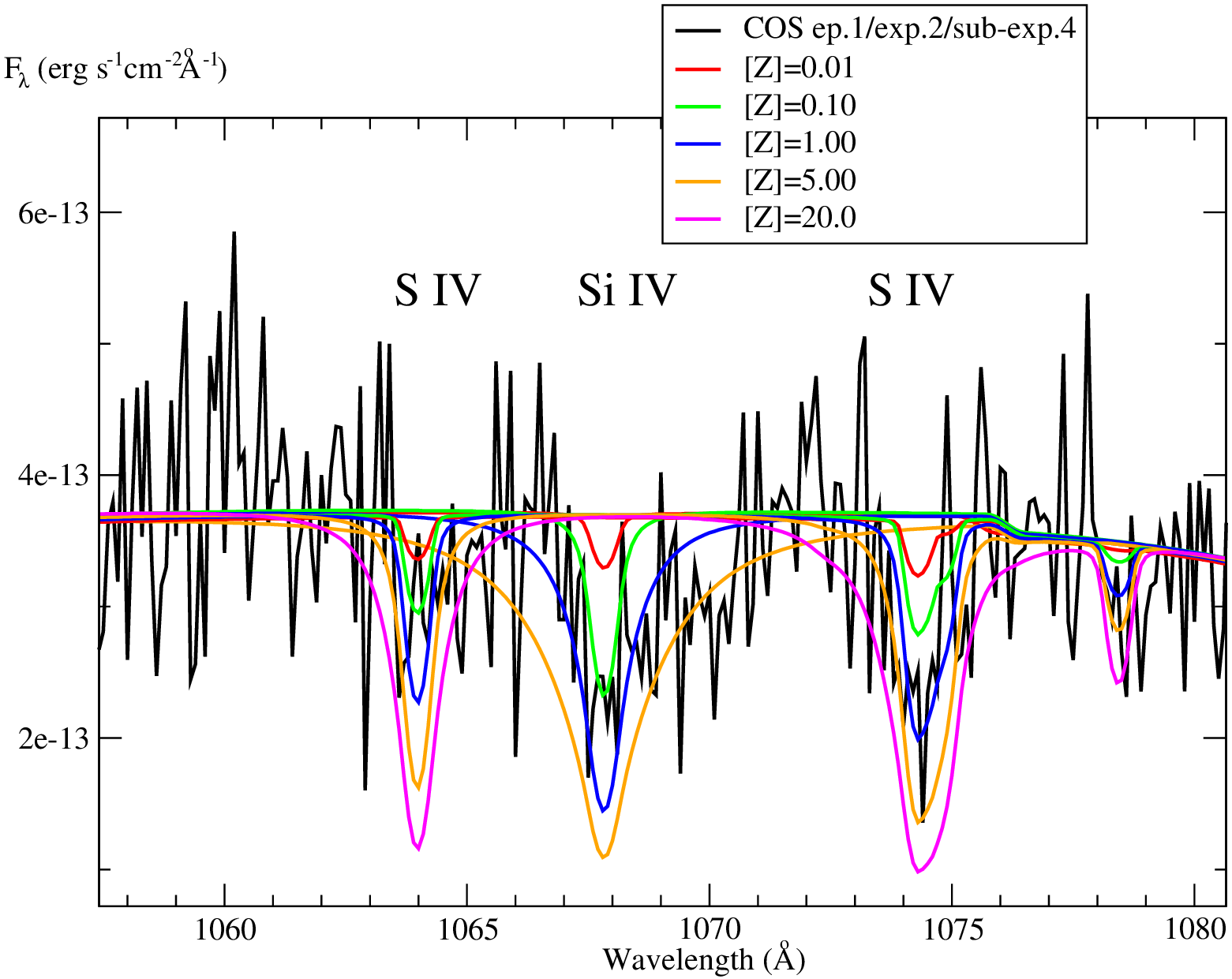}{0.30\textwidth}{(g)}
          \fig{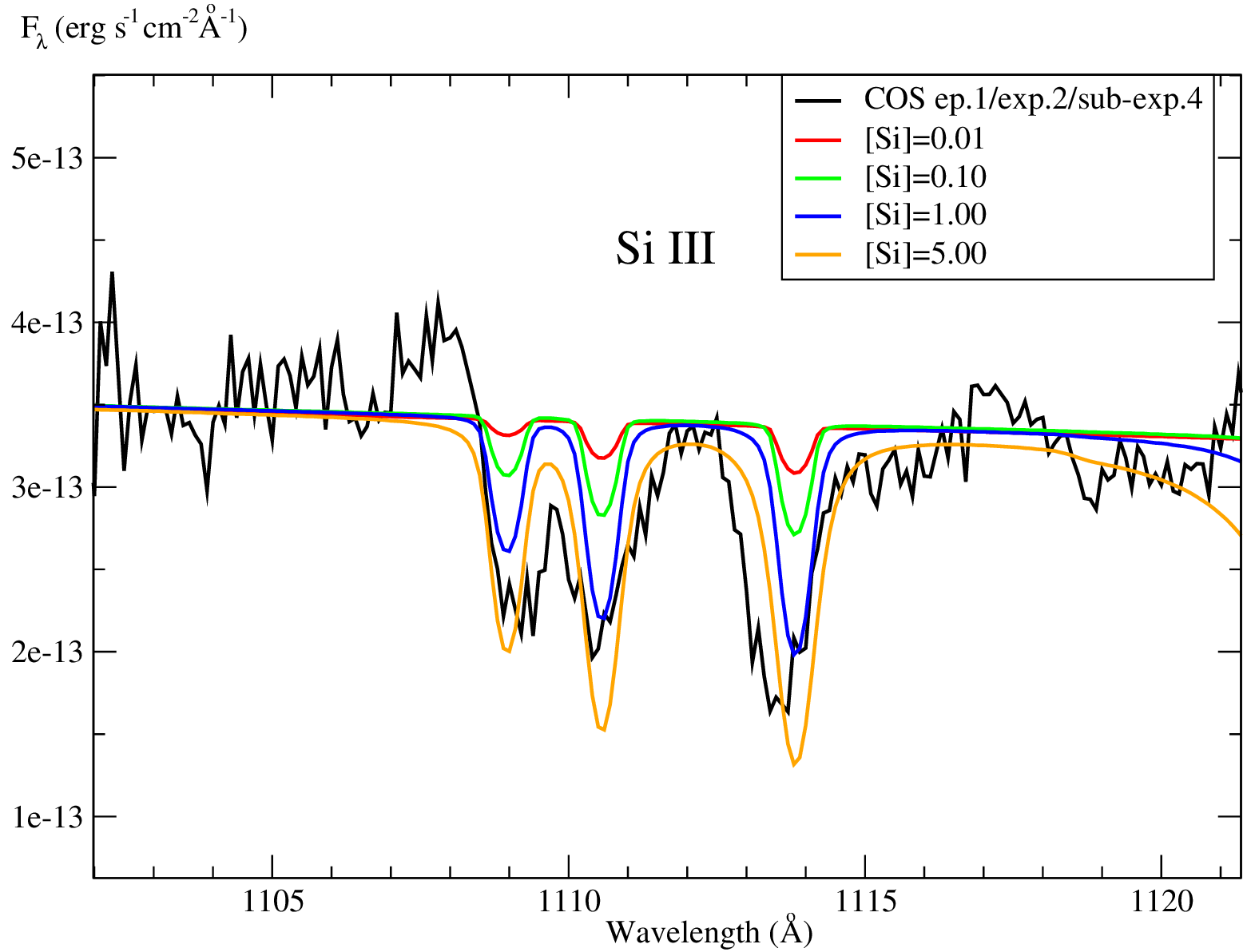}{0.30\textwidth}{(h)}
          \fig{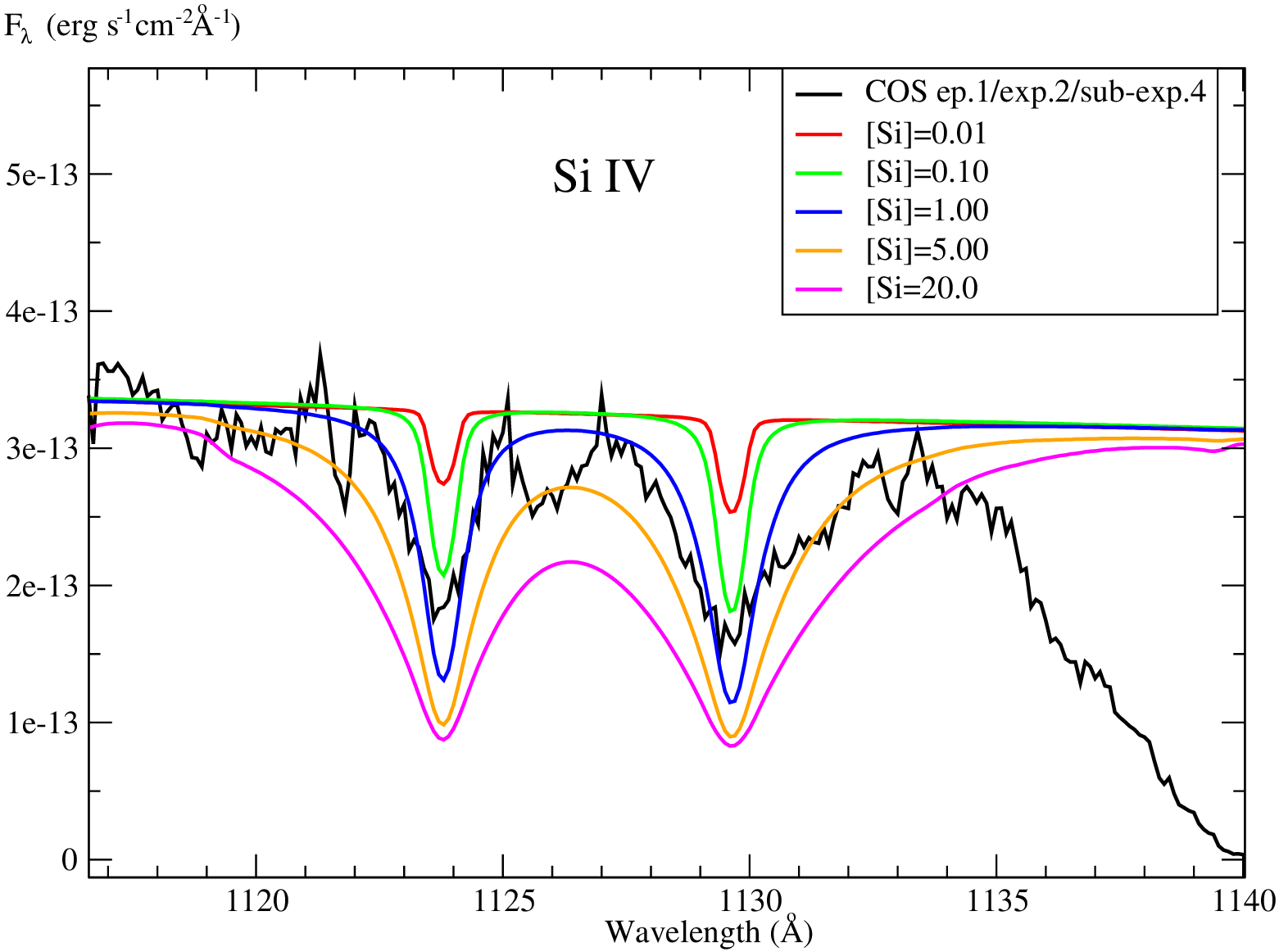}{0.30\textwidth}{(i)}
          }
\vspace{-2.7cm} 
\gridline{\fig{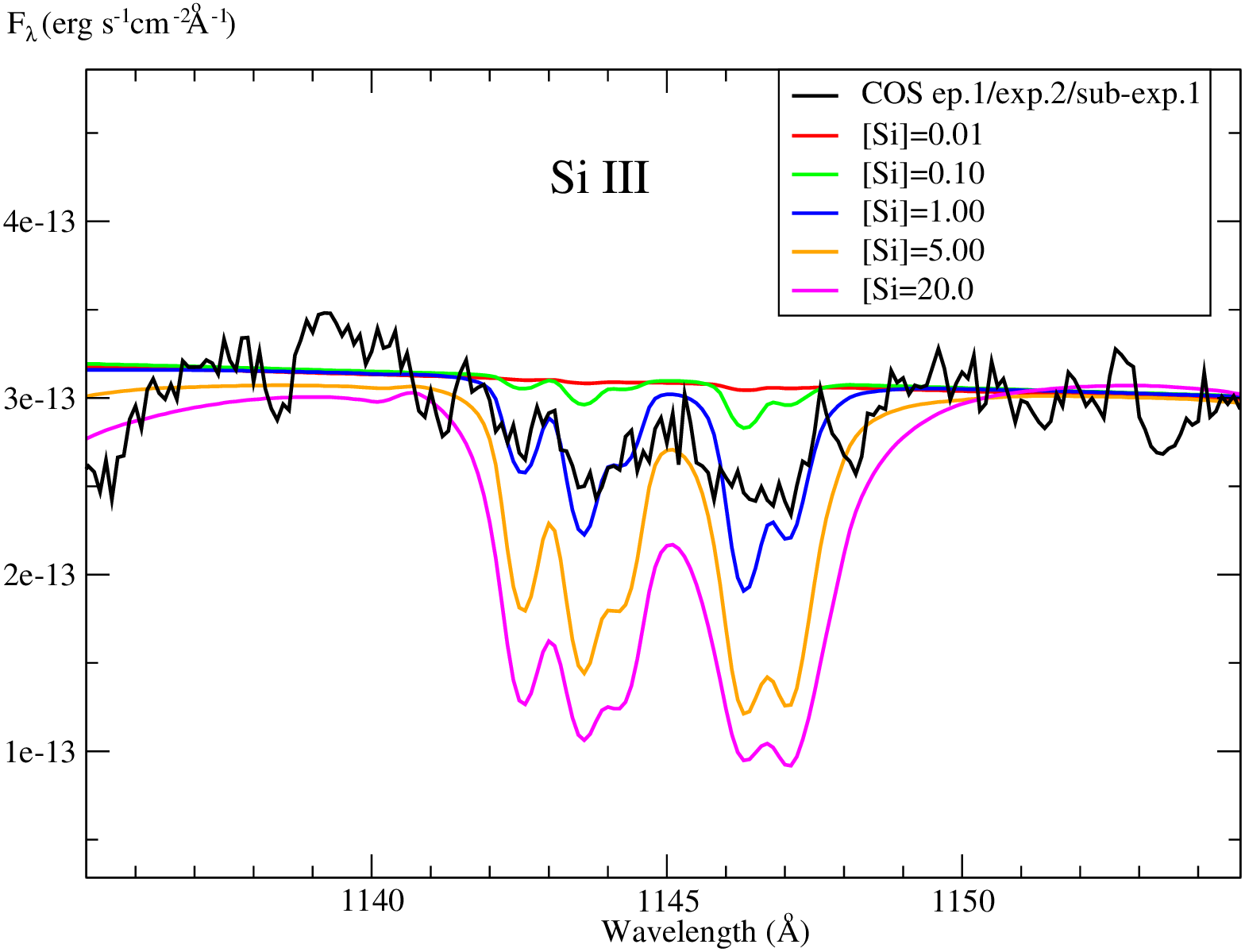}{0.30\textwidth}{(j)}
          \fig{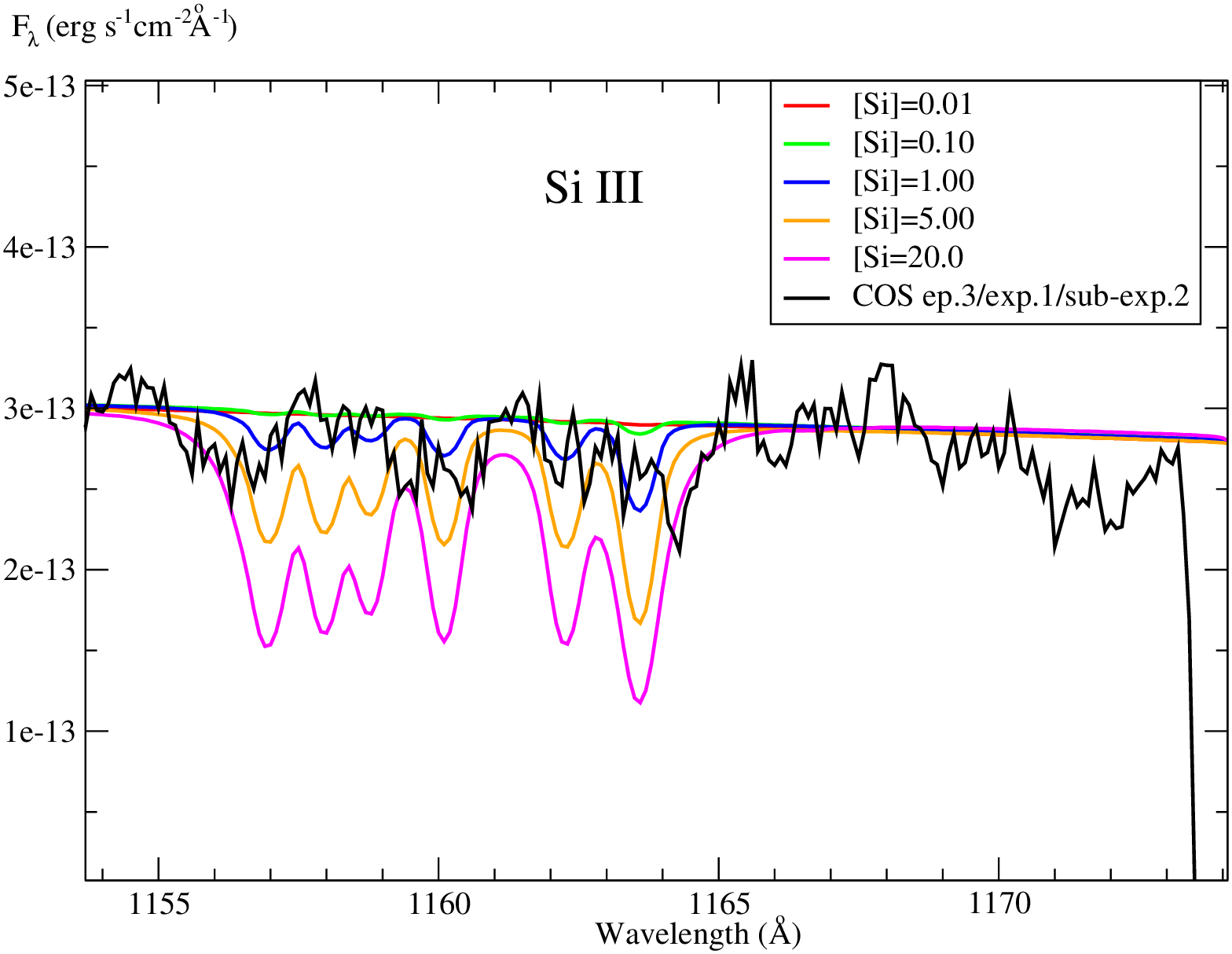}{0.30\textwidth}{(k)}
          \fig{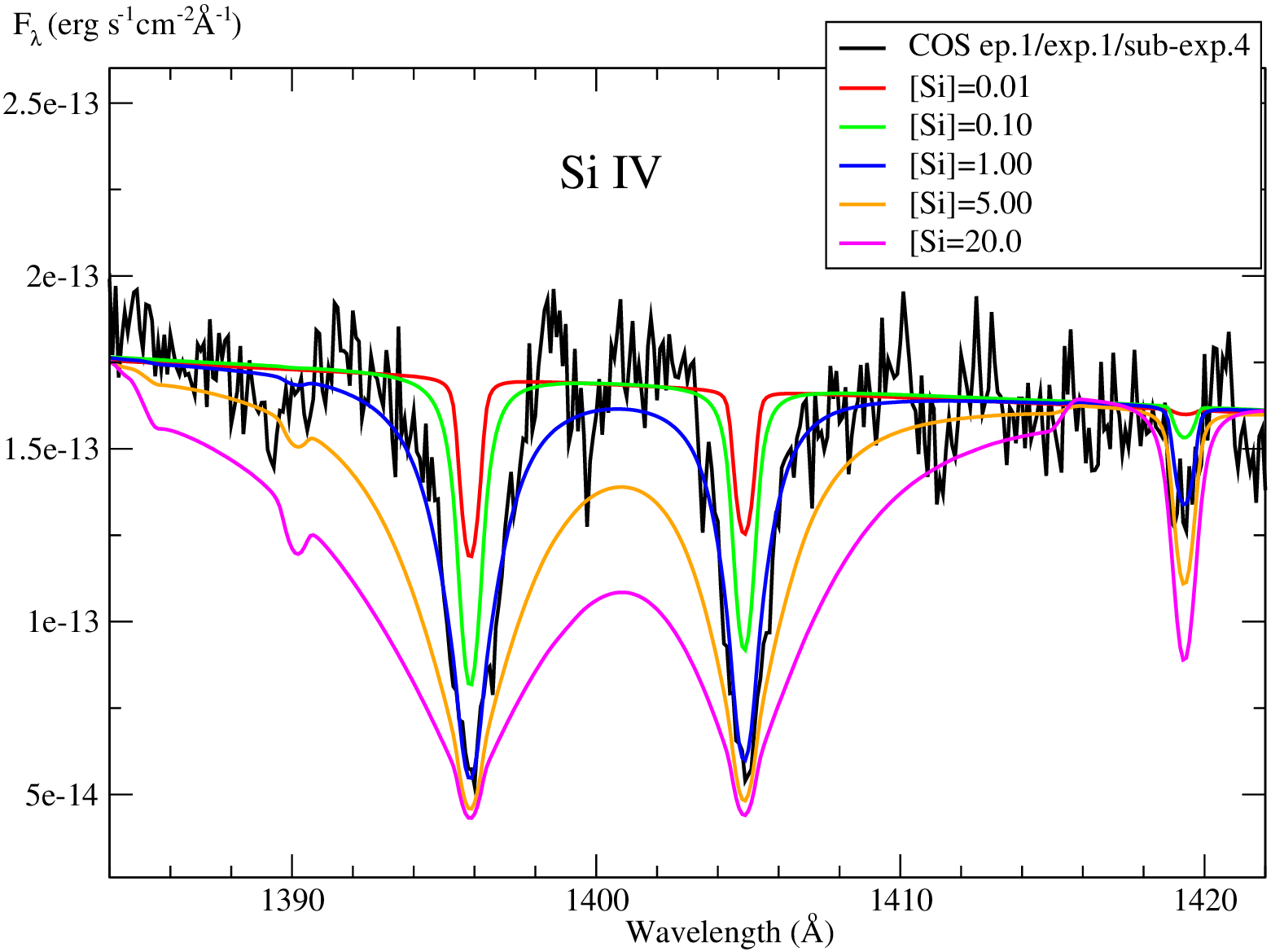}{0.30\textwidth}{(l)}
          }
\caption{
Abundance fits for the chemical elements listed in Table 4. 
In each graph we display an absorption feature at a given wavelength 
for a specific ionized species from a single sub-exposure 
spectrum (solid black line) together with the theoretical models
(color coded) as indicated in each panel. This figure qualitatively illustrates 
how the abundance for each ion listed in Table 4 was assessed.
See text for details.  
\label{fit} 
}
\end{figure*}

The fitting did not reveal abundances for O, P, Mg, Na or Ne (there were no
matching lines). 
In Table 4 we list the abundances we found for 
C, N, S, Si, and Al, 
and in Fig.\ref{fit} we illustrate the abundance fits for these 
ions. Some spectral regions (e.g. N\,{\sc III} $\sim$990, Fig.\ref{fit}(a)) 
are more noisy than others (e.g. Si\,{\sc iv} 1400, Fig.\ref{fit}(l)), 
and some line complexes (e.g. Si\,{\sc iii} 1110, Fig.\ref{fit}(h)) 
exhibit different line shifts within the complex. 
If one considers only the vicinity
of a given line (or complex), then in some cases the continuum 
is difficult to identify (e.g. Figs.\ref{fit}(a),(d)).   
For that reason the abundances were checked visually for each 
line in each sub-exposure. For the N\,{\sc iii} ($\sim$990) line
(Fig.\ref{fit}(a)), in spite of the fact that the
region is very noisy, the data do agree with a very large nitrogen abundance
[N]=20$\times$ solar or larger.  
For  the N\,{\sc iii} ($\sim$1085) absorption feature
(Fig.\ref{fit}, panel b), the depth of the 
line barely changes from [N]=0.01 (solid red line) to [N]=20.0 
(solid green line), and 
consequently one cannot rely on this feature to assess the nitrogen
abundance. In addition, 
the exact wavelength of the line is also difficult to assess:  
a [N]=1 model red-shifted by more than 1~\AA\ is also shown
for comparison (solid blue line) and it fits only the right wing of the 
line, while the two other models fit only the left wing of the line.  
As to the the N\,{\sc iii} (1184) absorption feature (panel c), 
it seems to be more reliable, however, the fitting of the adjacent 
C\,{\sc iii} (1175) line complex brings up the question of how to assess the
continuum flux level. 
For the nitrogen abundance derived from the N\,{\sc iii} (1184) absorption 
feature, as well as for most of the other abundances derived here, 
we used the continuum flux level obtained from fitting the entire
theoretical spectrum to the observed spectrum.  
For the C\,{\sc iii} (1175) absorption feature (panel d), we 
found that the left wing (including two additional smaller carbon
absorption lines to its left) and the right wing cannot be fitted
simultaneously. The fitting of the left side of the feature 
(as shown in panel d, giving [C]$\approx$0.4) yields a
higher continuum flux level than derived from the global fit.
The fitting of the right side of the feature, while giving the
same continuum flux level as derived from the global fit, does
not agree at all with the left side of the feature and produces
a line that is far too narrow ([C]=0.01).  
This could be due to detector effects near the edge of the spectral
segment, where the flux seems to drop. 
This wavelength region is only covered for one of the four
detector positions used with COS, i.e. only one in four sub-exposures,
we therefore decided not to rely on this carbon absorption feature,
nor on the N\,{\sc iii} 1184 feature. 
For the carbon abundance, we are left with the C\,{\sc iv} doublet
(1400, shown in panel e) which is covered in all the sub-exposures.  
Here the abundance is assessed mainly by fitting the wings of the 
absorption feature. The aluminum abundance is derived by fitting the
Al\,{\sc iii} doublet near 1860~\AA\ , as shown in panel f. For 
the sub-exposure displayed, the aluminum abundance is in excess of 
20$\times$ solar. The sulfur abundance was derived by fitting
the S\,{\sc iv} (1063 \& 1073) lines (shown in panel g). This
region of the spectral segment is very noisy and we found that the fitting of 
the 1063 line, which is to be narrower than the 1073 line,
gave a lower abundance than the 1073 line, possibly due to the noise.
Nevertheless, the sulphur abundance listed in Table 4 is an average
taken from fitting the two lines. Silicon abundances were derived 
from fitting 5 absorption line complexes. The Si\,{\sc iii} ($\sim$1112),
displayed in panel h, shows another example of a complex with lines
that are shifted relative to one another. The Si\,{\sc iv} ($\sim$1126)
displayed in panel i, shows how the right side of the absorption
feature (at $\lambda > 1132$~\AA ) has a reduced flux in the vicinity of the edge
of the spectral segment (followed by a gradual drop to zero at 1140~\AA ).  
The Si\,{\sc iii} (1143) absorption feature (panel j) is not very
pronounced or deep, but still provides a relatively good assessment
of the silicon abundance, contrary to the Si\,{\sc iii} (1160) absorption
feature (panel h) which is unreliable as it does not match the expected lines.
The Si\,{\sc iv} (1400) doublet (panel l) provides a more reliable
assessement of the silicon abundance, in particular with the 
additional small line seen around 1419~\AA\ .  
Accordingly, in Table 4, we marked all the absorption features
and abundances that are unreliable with an asterix. 

\begin{figure*}[t!] 
\vspace{-3.5cm} 
\gridline{\fig{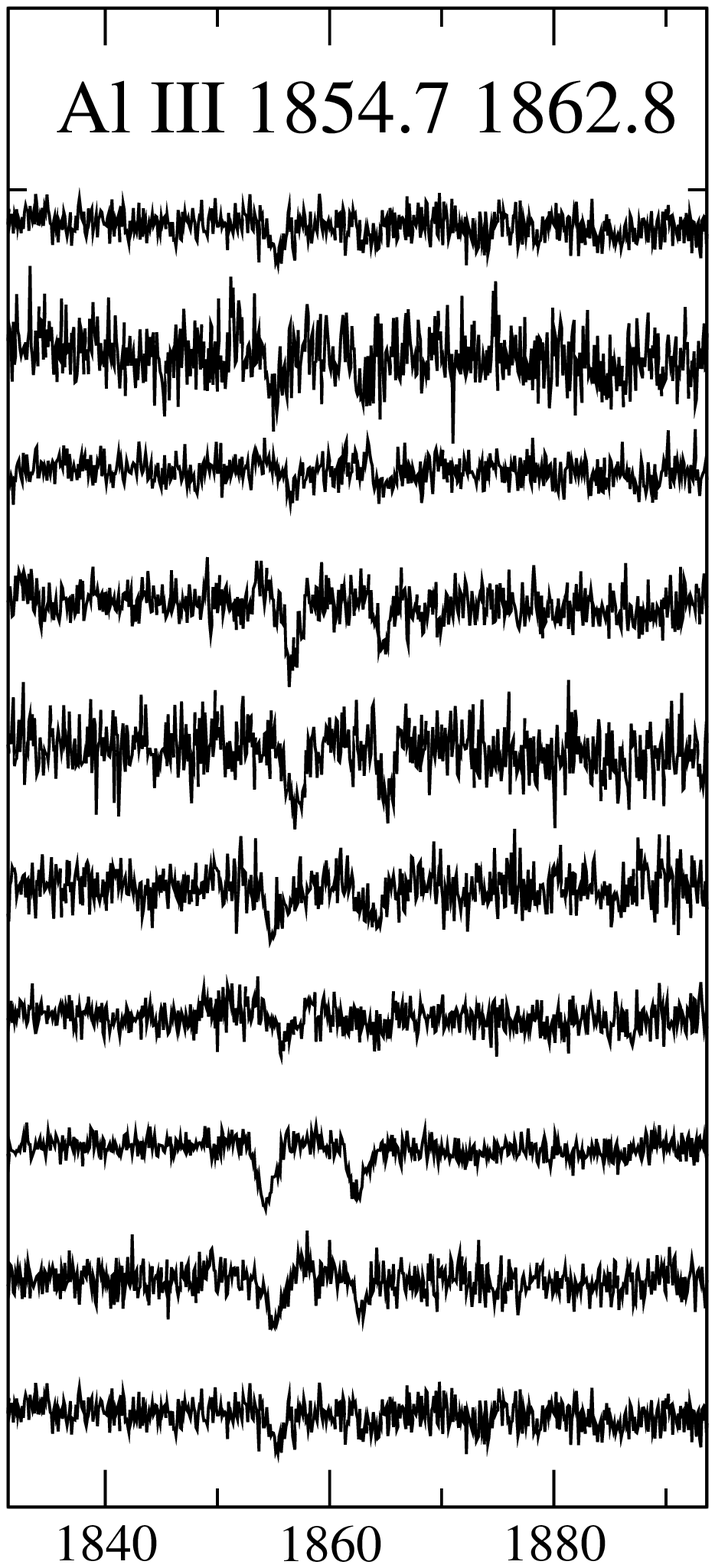}{0.45\textwidth}{Wavelength (\AA)}
          \fig{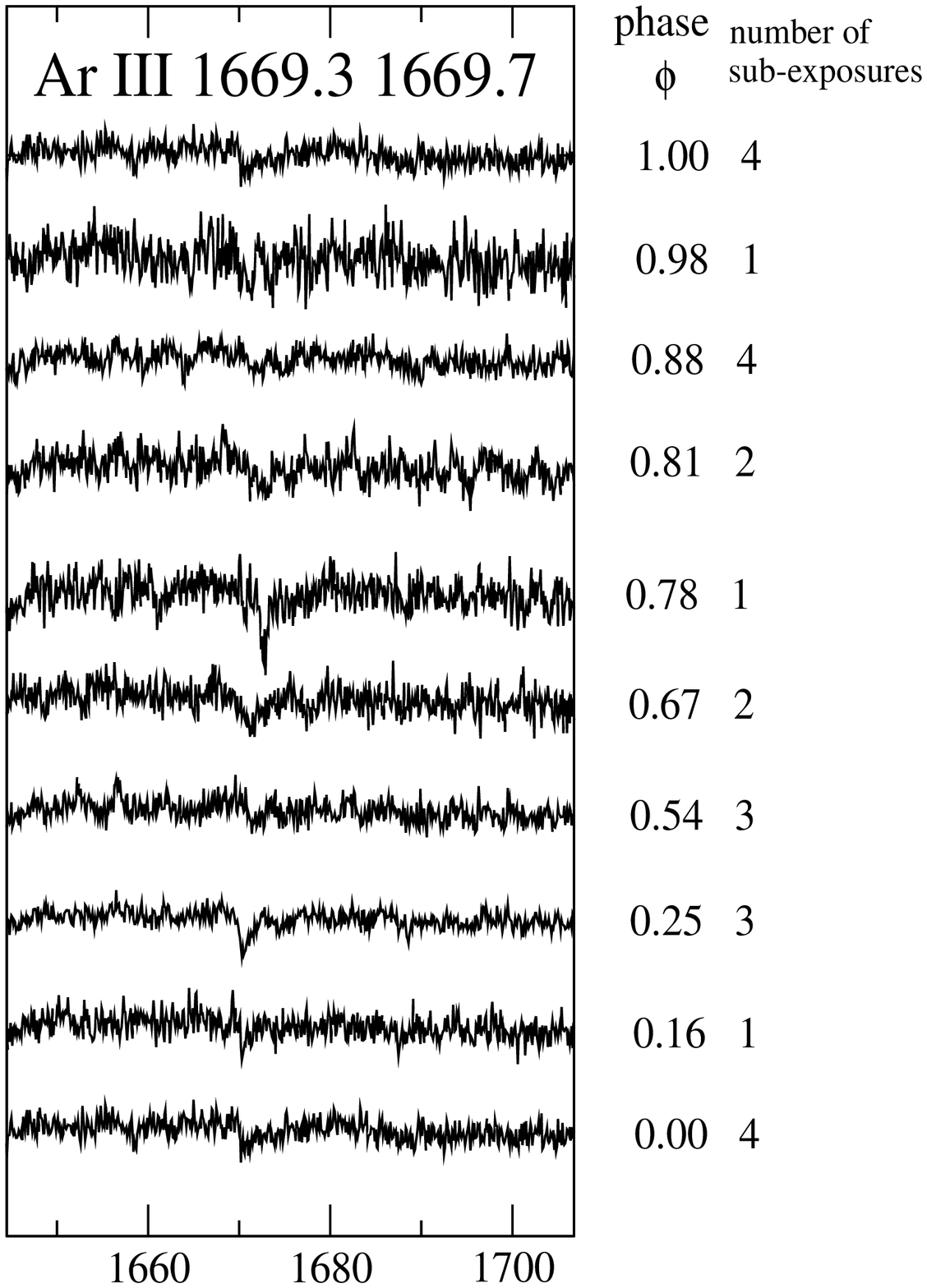}{0.45\textwidth}{Wavelength (\AA)}
          }
\vspace{-2.5cm} 
\gridline{\fig{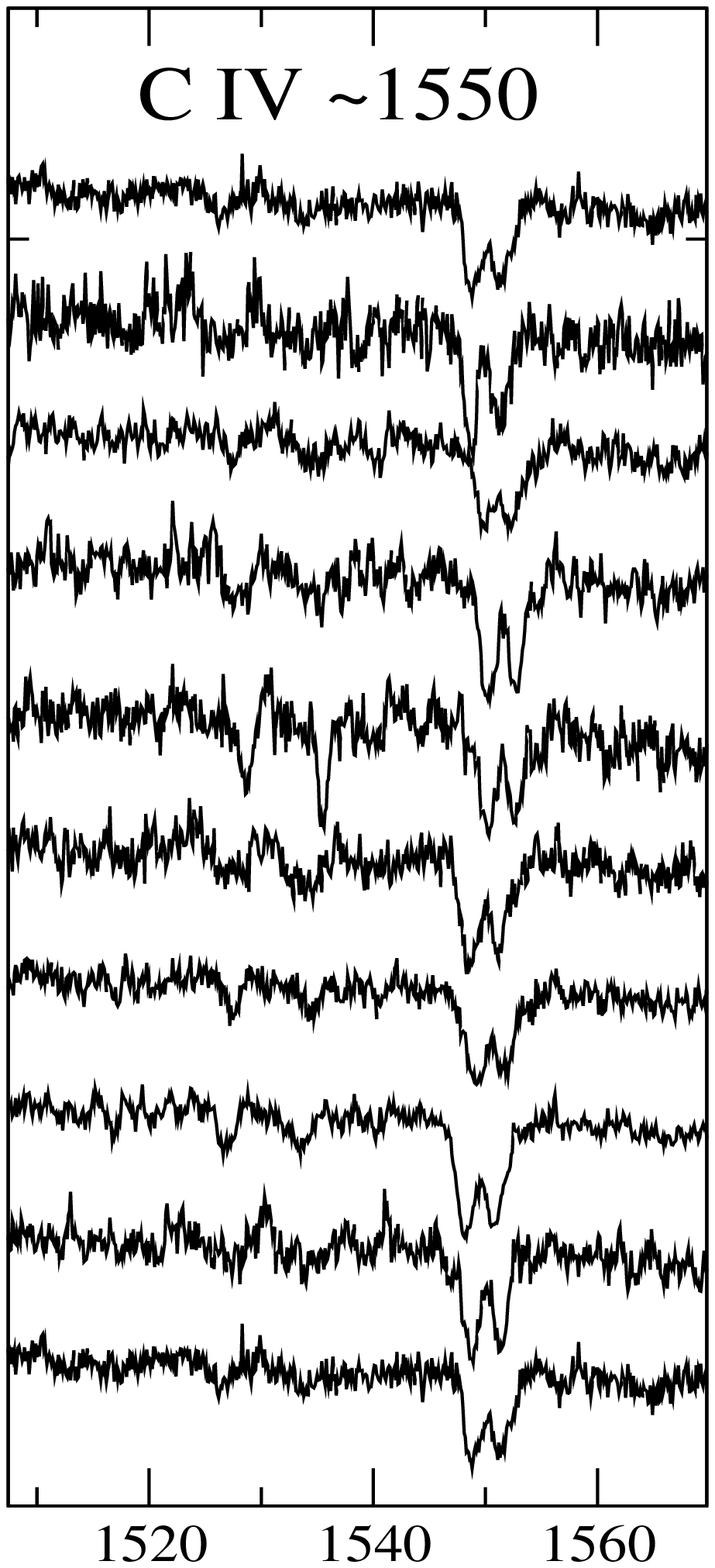}{0.45\textwidth}{Wavelength (\AA)}
          \fig{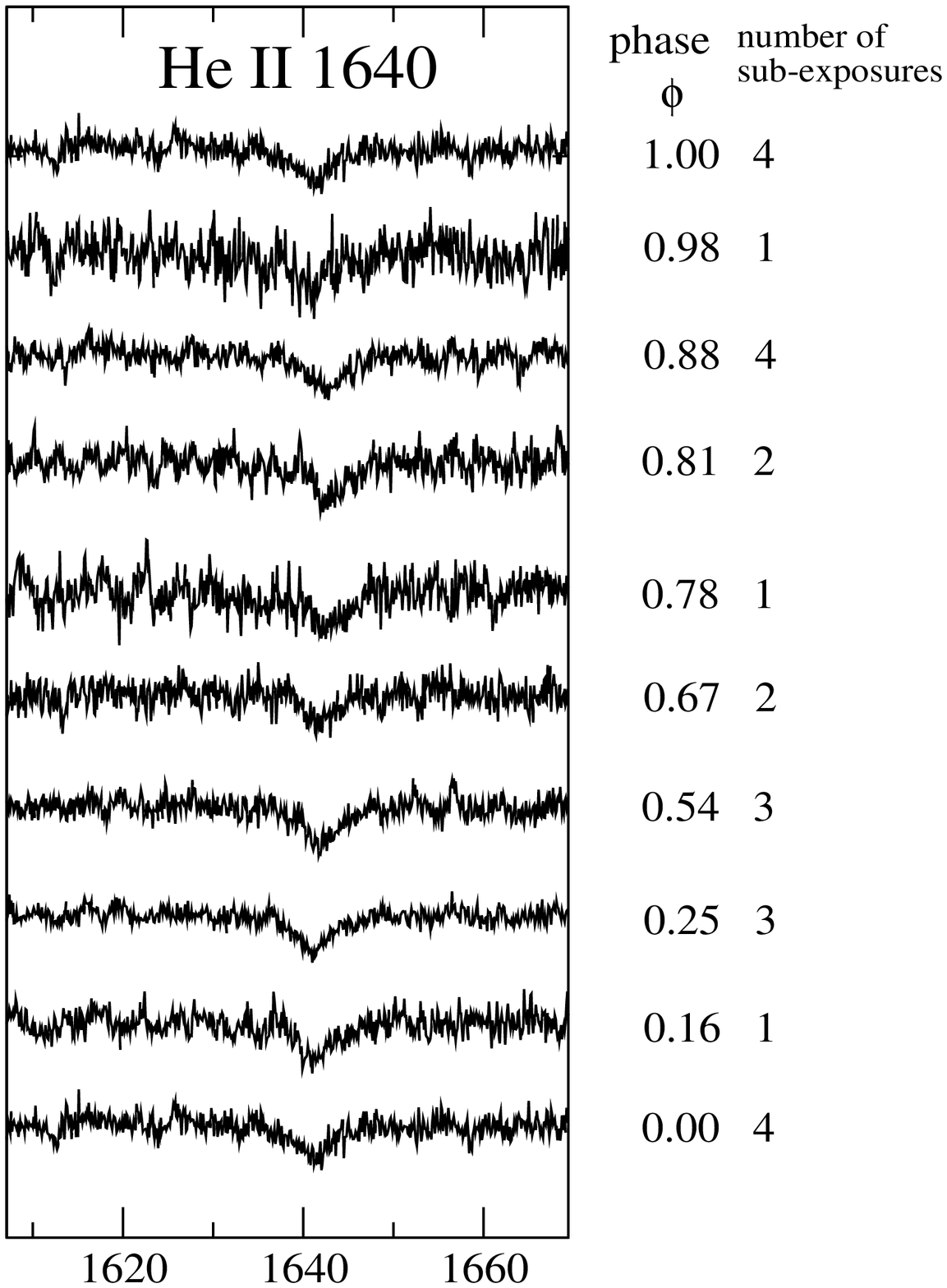}{0.45\textwidth}{Wavelength (\AA)}
          }
\caption{
The variation of the strength of the absorption lines as
a function of the orbital phase. The spectra were generated by
co-adding a number of sub-exposures 
in the immediate vicity of the orbital phase 
as shown on the right. 
The aluminum (Al\,{\sc iii} 1854.7 \& 1862.8) 
and argon (Ar\,{\sc iii} $\sim$1669.5) 
absorption lines (upper panels) are
stronger around phase $\phi \sim$0.25 and 0.75, and are much 
weaker (almost unoticable) around phases $\phi \sim$0.0 and 0.5. 
The carbon (C\,{\sc iv} $\sim$1550) and helium (He\,{\sc ii} 1640)
lines (lower panels) do now show such a strong phase variability. 
Note the strengthening of the Si\,{\sc ii} (1526.7 \& 1533.4) lines
around phase $\phi \sim 0.25$ and especially at phase 
$\phi =0.78$ 
(to the left of the C\,{\sc iv} lines in lower left panel).  
The spectrum at phase 1.00 (0.00) was obtained by combining the 
sub-exposure spectra from phases 0.96, 0.99, 0.00, and 0.03; 
the spectrum at phase 0.88 from phases 0.85, 0.86, 0.88, and 0.91; 
the spectrum at phase 0.81 from phases 0.80 and 0.81; 
the spectrum at phase 0.67 from phases 0.66 and 0.69; 
the spectrum at phase 0.54 from phases 0.51, 0.54, and 0.57; 
and the spectrum at phase 0.25 from phases 0.21, 0.25, and 0.30. 
\label{linesphasevar} 
}
\end{figure*}

For argon 
(Ar\,{\sc iii} $\sim$1669.5, see Fig.\ref{linesphasevar}),  we found
that the abundance would have to be 100$\times$ solar or larger.  
It is possible the $\sim$1669 absorption
features belongs to a different element which we could not identify.  

\begin{figure}[h] 
\vspace{-5.cm} 
\plotone{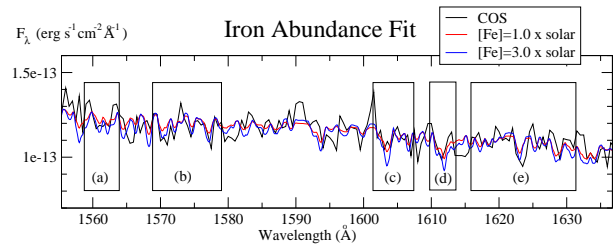}         
\vspace{-0.5cm} 
\caption{
The iron abundance, $[ {\rm Fe} ] = 2 \pm 1 \times$ solar, was assessed
by fitting the sawtooth pattern in the $\sim 1400-1600$~\AA\ 
region of the spectrum with a synthetic WD stellar model spectrum
(solid black line).
The model has a temperature of 41,500~K, a projected rotational velocity of 
150~km~s$^{-1}$, hydrogen and helium solar abundances, but no metal 
except for iron. The  iron was increased gradually until
the sawtooth pattern could be fit in some regions of the spectrum.
As an example we display some of these regions here 
in the rectangular boxes. From a careful examination we found that 
lines could be fitted for [Fe]=1 $\times$ solar (e.g. box d) 
to  [Fe]=3 $\times$ solar (e.g. box b).  
The  COS spectrum displayed here (solid red line) 
is the first sub-exposure of the second orbit obtained at epoch \# 1
(phase $\phi=0.81$). Similar fits were obtained for all sub-exposures.  
\label{iron} 
}
\end{figure}

The iron abundance was derived by fitting the sawtooth pattern 
in the wavelength range $\sim 1500-1600$~\AA , as shown in 
Fig.\ref{iron}. 

Overall we found that carbon and sulfur are both consistent 
with sub-solar to solar abundances; nitrogen, aluminum and argon 
have  all supra-solar abundances. 
Silicon  is sub-solar except 
for the Si\,{\sc iii} ($\sim 1112$~\AA ) and Si\,{\sc iv} (1400~\AA )
lines that match about solar abundances, confirming the claim of 
\citet{lon06} that the    
Si\,{\sc iii} lines at $\sim$1110~\AA\  require abundances 
that are much larger than required at 1140 and 1155~\AA .
We note, however, that the silicon lines at $\sim 1112$~\AA\ 
can be fit with a lower temperature, namely $T_{\rm eff}=25-35,000$~K.
We will come back to this in section 3.4.     
The supra-solar nitrogen abundance with the sub-solar carbon
abundance further confirms the anomalously high N/C ratio
observed in previous spectra of U Gem in quiescence
that has been attributed to CNO processing \citep{sio98,lon06}. 

At solar abundance, the He\,{\sc ii} 1640 absorption feature 
needs a higher rotational
velocity of 200-250~km~s$^{-1}$ in  some of the sub-exposure spectra.
We also found that
the Si\,{\sc iv} doublet ($\sim 1400$~\AA ) 
appears to be off (red-shifted) by about one Angstrom compared to other lines, 
as noted by \citet{lon06} who 
suggested  that they not always form in the stellar photosphere.  
Some of the absorption lines appear to be transient (see next subsection). 
We also found that the width, shape, shift
and depth of many lines are not uniform and vary, even in a single
sub-exposure, indicating that the lines do not form in the same medium.    
At a first glance (e.g. Table 4), the abundances  
do not seem to change as a function of the time since outburst. 
We, therefore, decided to analyze the absorption lines as a function of the
orbital phase.

\clearpage

\subsection{{\bf Orbital Phase Variability of the Absorption Lines}} 
We use the ephemeris of U Gem as given by \citet{ech07} to find the 
phase at which each sub-exposure was obtained, where we take the phase
$\phi=0$ as the secondary conjunction. The phase of the start of 
each exposure is indicated in Fig.\ref{si4doublet} 
as well as in the last column 
in Table 4 to within an accuracy of 0.01 ($\sim 150$~s).  

By visual inspection of the sub-exposure spectra, we 
found that many absorption lines appear stronger/deeper
at some orbital phases, regardless of the epoch 
at which the spectra were taken. We, therefore, decided to examine the 
sub-exposure spectra in order of increasing orbital phase from 0.0 to 1.0 and
to combine them near the same orbital phases when possible.   
Following this procedure we generated 9 spectra at orbital phases
0.00 (made of 4 sub-exposures), 0.16 (1 sub-exposure), 0.25 (3), 
0.54 (3), 0.67 (2), 0.78 (1), 0.81 (2), 0.88 (4) 0.98 (1), and 
1.00 (=0.00). We found that the Al\,{\sc iii} (1854.7, 1862.8), 
Ar\,{\sc iii} ($\sim$1669.5), and Si\,{\sc ii} (1526.7, 1533.4)
absorption lines are particularly sharp and deep at orbital phases
$\phi=0.25$ and 0.78, and are much weaker (almost absent) 
around orbital phases $\phi \sim 0.54$ and 0.00 (1.00) 
(see Fig.\ref{linesphasevar}). 
Al\,{\sc iii} appears to be stronger at phases 0.78 \& 0.81 and 
0.16 \& 0.25, namely over a wider range of phases.

\begin{figure}[h!] 
\vspace{-2.cm}
\plotone{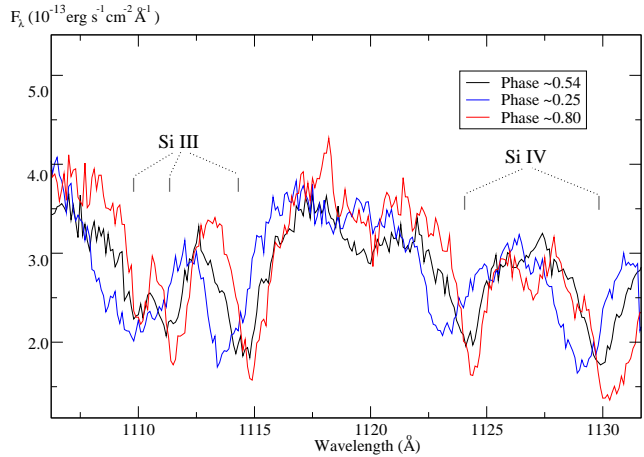} 
\vspace{-1.0cm}
\caption{
The silicon lines in the wavelength range 1100-1130~\AA\ 
are deeper around phase 0.80, and shallower around
phase 0.54 (and 1.00 - not shown for clarity). The increase 
in the depth of the absorption lines indicates the presence of
absorbing material in front of the WD at phases $\sim 0.25$ and
$\sim 0.80$. The spectrum around phase 0.54 was created by combining
together the sub-exposure spectra from phases 0.51, 0.53, 0.54 and 0.57, 
the phase 0.25 spectrum was obtained from combining phases 0.21, 0.24, 0.25 
and 0.30, and that of phase 0.80 was from combining phases 0.78, 0.80 
and 0.81.    
\label{siphasevar} 
}
\end{figure}

Most of the other
metal absorption lines (especially the silicon lines in the range
$\sim 1110$ to $\sim 1130$~\AA ) seem to exhibit some change 
with the orbital phase, where the depth of the absorption lines 
decreases from phase $\sim$0.80 (deepest), to phase $\sim$0.25 (intermediate),
and phases $\sim$0.0 and $\sim$0.54 (shallowest, 
see Fig.\ref{siphasevar}).
The Si\,{\sc iv} (1400), and C\,{\sc iv} (1550) absorption lines 
are only slightly deeper around phase 0.25 and 0.80, while the 
He\,{\sc ii} (1640) absorption line does not exhibit such a variation.   
Due to the wavelength coverage of the sub-exposure spectra
we could not carry out the same analysis for the 
the C\,{\sc iii} (1175) and N\,{\sc iii} (1183) absorption lines. 
In addition, the low S/N below 1100~\AA\ 
prevented us from obtaining robust results for absorption lines
in that wavelength region.  

In spite of the fact that the sub-exposure spectra were obtained at
4 different epochs, the results show a consistent  
increase in the depth of the absorption lines at orbital
phases 0.16-0.30 and 0.67-0.81, indicating the
presence of an absorbing material with high metal content
(i.e. Si, Al, Ar) in front of the WD  at those phases.  
This is in complete agreement with the results of \citet{fro01,lon06} 
who found an increase in depths of the lines 
and the sudden appearance of low-ionization lines at orbital phase
between 0.53 and 0.85.
These are  the same orbital phases
at which the X-ray and EUV light-curves dip, as seen in U Gem in both outburst
and quiescence \citep{mas88,lon96,szk96}. The increase in the depth 
of the absorption lines and the EUV/X-ray light curve dips have
possibly the same origin, i.e. a (rim) disk bulge with mass stream
overflow material located at large disk radii to agree with the observed
low radial velocity \citep{fro01,lon06}.  
The increase in the depth of the lines around phase 0.16-0.30 
can be attributed to the stream material overflowing the
disk rim and bouncing off the disk surface near phase 0.5 
where it is then redirected toward phase $\sim 0.10-0.30$
\citep{kun01}.  

\begin{figure}[h!] 
\vspace{-1.5cm}
\plotone{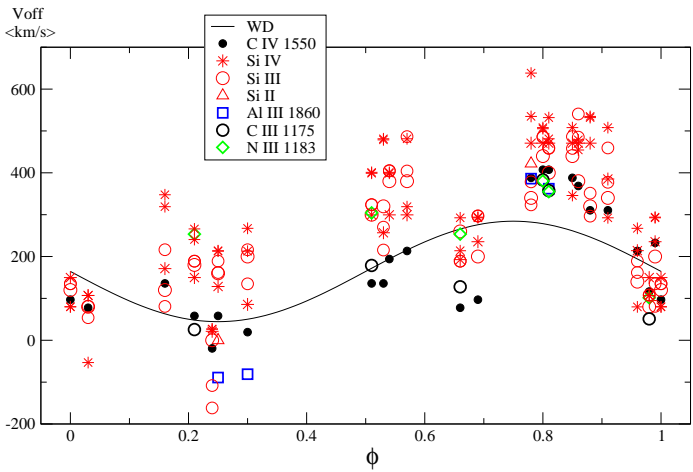}  
\vspace{-0.7cm}
\caption{ 
Absorption lines velocity shift (in km~s$^{-1}$) in the COS spectrum of 
U Gem as a function of the orbital phase $\phi$. The velocity offsets were
measured from the rest wavelengths for the 23 sub-exposures using 
a synthetic WD photosphere spectrum. The solid black line is the orbital
motion of the WD assuming an orbital velocity amplitude of 
$\sim$122~km~s$^{-1}$ \citep{lon06},     
a WD gravitational redshift of $\sim$80.4~km~s$^{-1}$
\citep{sio98}, 
and a proper motion radial velocity of $\sim$84~km~s$^{-1}$
for the U Gem \citep{wad81}. The carbon lines (black circles \& dots) follow 
more closely the WD orbital velocity. The Si\,{\sc iv} 
($\lambda \lambda$ 1122.5, 1128.3, $\sim$1400)  
lines (red stars) 
exhibit the largest offset from the WD orbital velocity, 
followed by Si\,{\sc iii} ($\lambda \lambda \sim$1113, $\sim$1144, 1500)
lines (red circles)  \& Si\,{\sc ii} ($\lambda \lambda$1530)
lines (red triangles).  Only a small number of data points could be 
obtained for nitrogen (green diamonds) and aluminum (blue squares).  
\label{linesphasevarvel} 
} 
\end{figure}

In order to try and further identify the location of the formation of
the absorption lines in the COS spectrum of U Gem, we measured 
the velocity shift of the lines in each sub-exposure spectra. 
We match each line individually with the corresponding line in a synthetic 
stellar atmosphere spectrum generated with TLUSTY/SYNSPEC, and we 
(red- or blue-) shifted the spectrum until the observed line became
superposed to the
theoretical line. In this manner we obtained a spectral shift accurate
within $\sim$0.1~\AA\ , which we then translated into a radial velocity
shift. The lines for which the velocity was measured were the
C\,{\sc iv} (1550), Si\,{\sc iv} (1400, 1122.5, 1128.3), 
Si\,{\sc iii} (1113, 1144, 1500), and Si\,{\sc ii} (1530).  
For the C\,{\sc iii} (1175),  N\,{\sc iii} (1183), and Al\,{\sc iii} 
($\sim$1860) lines only a limited number of dataset were available.   
The velocity shift of the absorption lines is displayed 
as a function of the orbital phase in Fig.\ref{linesphasevarvel}, 
together with the expected
velocity shift of the WD, taking into account a 122~km~s$^{-1}$ orbital
velocity \citep{lon06}, a 80.4~km~s$^{-1}$ gravitational redshift
\citep{sio98}, and a recessional velocity of 84~km~s$^{-1}$ \citep{wad81}.
The first thing that is striking in Fig.\ref{linesphasevarvel} 
is the large scatter
observed at all phases, but this scatter has been observed before
\citep{fro01}. 
The carbon lines (in black) appear to follow the WD velocity more closely than 
the other lines, and the Si\,{\sc iv} lines seem to have the largest
velocity offset, followed closely by the Si\,{\sc iii} lines. 
\citet{lon06} also found the Si\,{\sc iv} doublet ($\sim 1400$~\AA ) 
to be red-shifted by about one Angstrom compared to the other lines, 
suggesting  that they not always form in the stellar photosphere.  
For the most
part the velocity offset of the silicon is positive, indicative of material
falling toward the WD. The average velocity offset is the largest (positive)
at phases 0.78-0.91, which is consitent with the location of the L1-stream
material overflowing the disk's edge and falling toward smaller radii.

\subsection{{\bf Orbital Phase Variability of the Continuum.}}  

\begin{figure}[h!] 
\vspace{-6.5cm} 
\gridline{
\fig{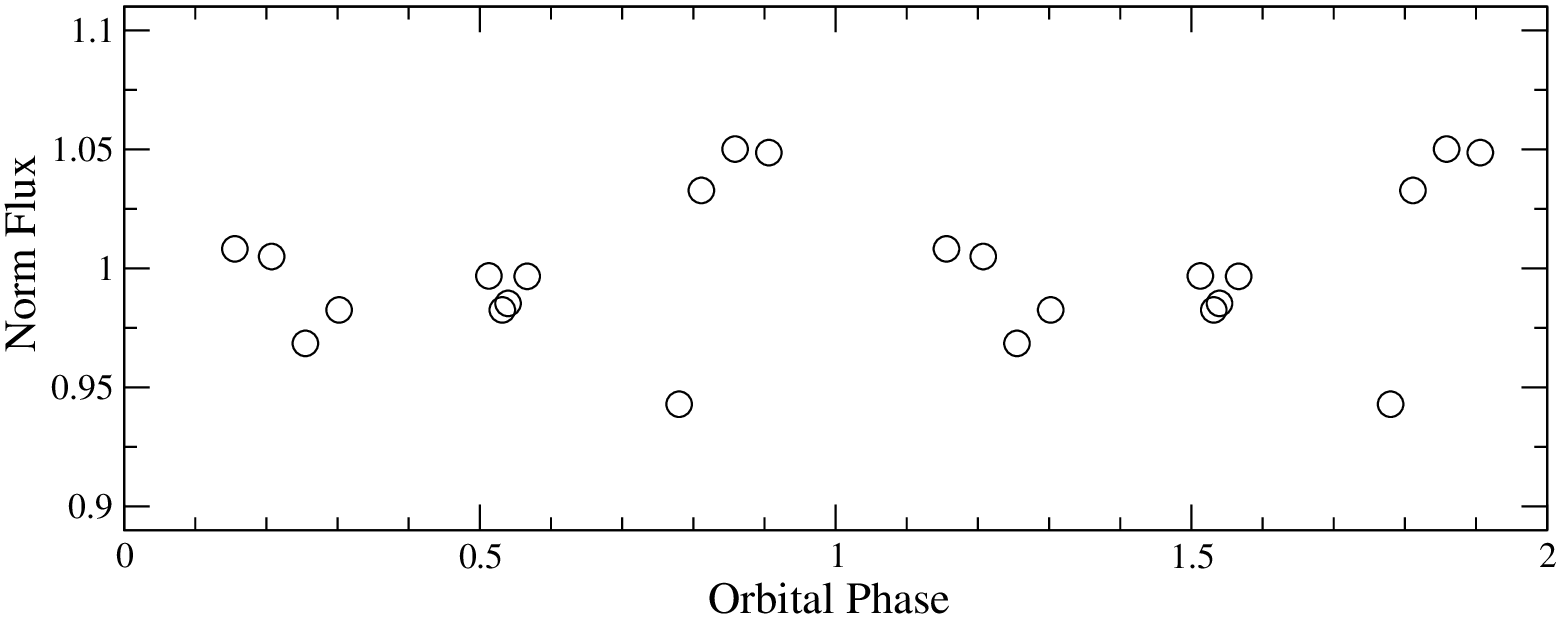}{0.45\textwidth}{}
} 
\vspace{-5.4cm} 
\gridline{
\fig{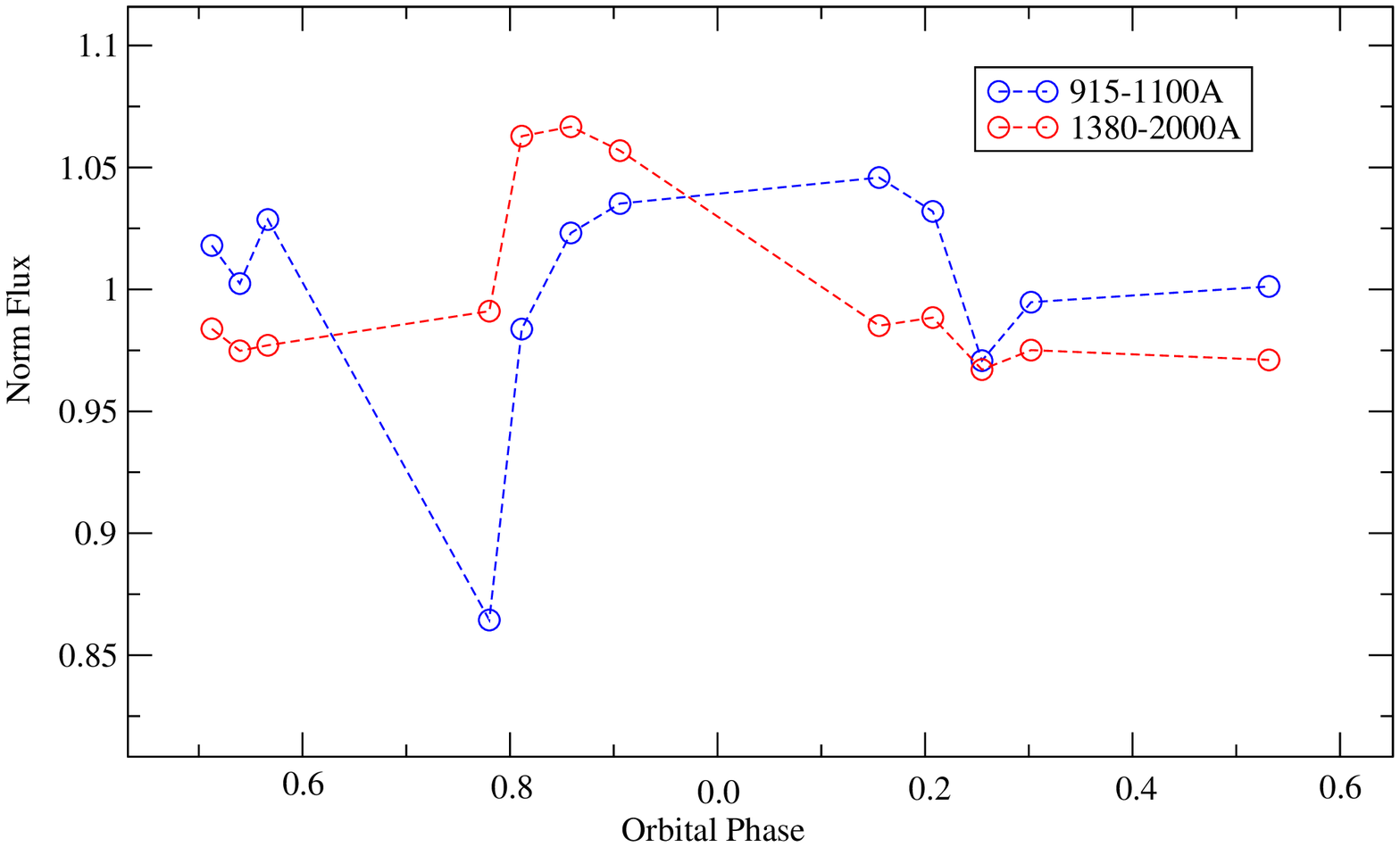}{0.40\textwidth}{}
}
\vspace{-0.5cm}  
\caption{
{\bf Top.} 
The UV lightcurve as a function of the binary phase generated by 
integrating the 12 sub-exposures from epoch \# 1 
over the spectral region 915-2000~\AA , 
from the {\it HST}/COS observation of
U Gem, obtained 15 days after the peak of a ``wide'' outburst. 
There is a 
dip at $\phi = 0.25$ and a second deeper dip at     
$\phi = 0.78$, and a maximum toward $\phi \sim 0.8-0.9$.  
The total flux (915-2000~\AA ) varies by only 
$\pm$5\%.
\label{fluxphasevara} 
}
\vspace{-0.3cm} 
\caption{
{\bf Bottom.} 
Normalized flux in the wavelength ranges 915-1100 \AA\ (in blue)
and 1380-2000 \AA\ (in red) as in Fig.\ref{fluxphasevara}. 
The two dips observed in Fig.\ref{fluxphasevara}
are an actual feature of the shorter
wavelengths only, while the maximum is more pronounced
in the longer wavelengths. 
\label{fluxphasevarb} 
}
\end{figure}

The 12 sub-exposures of the first epoch are covering a complete
binary orbit and where all obtained within that time frame. 
The sub-exposures obtained at epoch \# 2, 3 and 4 have a lower
flux since they were obtained as the system was declining.  
Consequently, 
in order to check the variability of the continuum in a self-consistent
manner, we used only the 12 sub-exposures from the first epoch.  
For each of the 12 sub-exposures we integrated the flux from 915~\AA\ 
to 2000~\AA\ and normalized it. The resulting COS FUV lightcurve  
is presented in Fig.\ref{fluxphasevara} 
and exhibits an orbital phase variability 
with a maximum amplitude of only $\pm$ 5\%, 
which is relatively small compared to +20\% and -30\% observed in
{\it FUSE} observations obtained in deep quiescence \citep{lon06}.
Even though the lightcurve is made of only 12 data points covering 
the entire binary orbit, we notice that the two lowest points are 
at $\phi=0.25$ (local minimum) and $\phi=0.78$ (global minimum), 
with maximum at phases $\phi = 0.81-0.91$.     

In order to make a better comparison with the {\it FUSE} and optical
lightcurves,
we generated two lightcurves, one covering the shorter wavelengths
915-1100~\AA\ , and the other one covering the 1380-2000~\AA\ region
(Fig.\ref{fluxphasevarb}).

The two dips observed in Fig.\ref{fluxphasevara}
appear mainly as features in the 
short wavelength lightcurve only (in blue in Fig.\ref{fluxphasevarb}). 
The two dips correspond to veiling of the WD by the same 
L1-stream material that has overflowed the
disk rim (near $\phi \sim 0.9$): 
first as it  overflows higher above the disk 
toward smaller radii near $\phi=0.78$,  
then as it is re-directed toward
phase $\phi \sim$0.25 after it has bounced off the disk surface  
near phase 0.5 \citep{kun01}. 
This is completely consistent with the
{\it FUSE} lightcurve revealing dips at orbtial phase 0.6-0.85
as well as at orbital phase 0.2-0.35. 
Note that in the {\it FUSE} lightcurve the flux below 970~\AA\ 
decreases substantially during the $\sim 0.7$ phase dip
\citep{lon06}.

The longer wavelength lightcurve (Fig.\ref{fluxphasevarb}) 
exhibits an increase of flux near phase 0.8-0.9.  
This is similar to the optical (blue and red) 
light curves revealing a strong hump  
at phase 0.8-0.9 due to the contribution of the hot spot facing the observer 
and before it is occulted by the secondary near phase 0.0 \citep{und06,ech07}. 
A secondary broad hump around phase 0.25 is also observed in the 
H$\alpha$ \& H$\beta$ light curves \citep{mar90,und06,ech07}.

\subsection{{\bf The White Dwarf Temperature}}

Since many absorption lines in the spectrum of U Gem  
are either forming in an absorbing medium in front of the WD, or 
are being affected by that absorbing medium, we decided to model the
WD spectrum using the continuum and not concentrate on the {\bf metal}  
absorption lines in the model fit. 
The continuum is essentially characterized by its slope and 
by the broad absorption features of the hydrogen Lyman and
helium Balmer series. The advantage of the COS spectrum is that it
covers the spectral range down to the Lyman limit (though the Ly$\alpha$
feature falls into the detector gap) and consequently allows us to derive the
WD temperature by fitting the Lyman series. While the continuum slope
can also be used to derive the WD temperature, it is not as sensitive
to the temperature as the Lyman series (for the WD temperatures considered
here). 
For our WD temperature analysis,  we combined the sub-exposure 
spectra by epoch to generate 4 spectra, one for each epoch. 
Since the first epoch has 12 sub-exposures, its spectrum has a much
higher S/N than the second, third and fourth exposure spectra. 
Also, since the lines have a `scattered' velocity broadening of 
about 300~km~s$^{-1}$ (see Fig.\ref{linesphasevarvel}; reaching 
almost  400~km~s$^{-1}$ at $\phi=0.25$), we shifted only the first epoch 
sub-exposure spectra by the WD orbital velocity before  combining them. 
As a results the absorption lines have a velocity broadening
larger than the known WD projected rotational velocity of 
100-150~km~s$^{-1}$.   
In order to approximately match the absorption lines we set up solar
abundances for all elements, except for nitrogen and aluminum that
were both set to $20 \times$ solar. 
We then modeled the 4 epoch spectra with synthetic WD (with $\log(g)=8.8$)
model spectra generated in the manner described at the very beginning 
of this section. \\ \\  

\subsubsection{The First Epoch.} 

As we fitted synthetic WD model spectra to the first epoch spectrum, 
we found that the slope of the COS continuum 
in the longer wavelength down to 2000~\AA\ agreed with a  
temperature as low as 
25,000~K which gives an extremely short distance (43pc) 
but did not provide enough flux below 1200~\AA\ 
and basically no flux below 1000~\AA\ . 
As we increased the temperature to 35,000~K, 
the slope of the theoretical continuum decreased very
slowly, while the short wavelength
region was still deficient in flux and the profile of the Lyman
series were not fitted properly. In the longer wavelength the slope
of the continuum agreed down to about $\lambda \sim $1800~\AA\ but 
the model was very slightly deficient in flux for $\lambda > 1800$~\AA .  

\begin{figure*}[t!] 
\vspace{-9.cm} 
\plotone{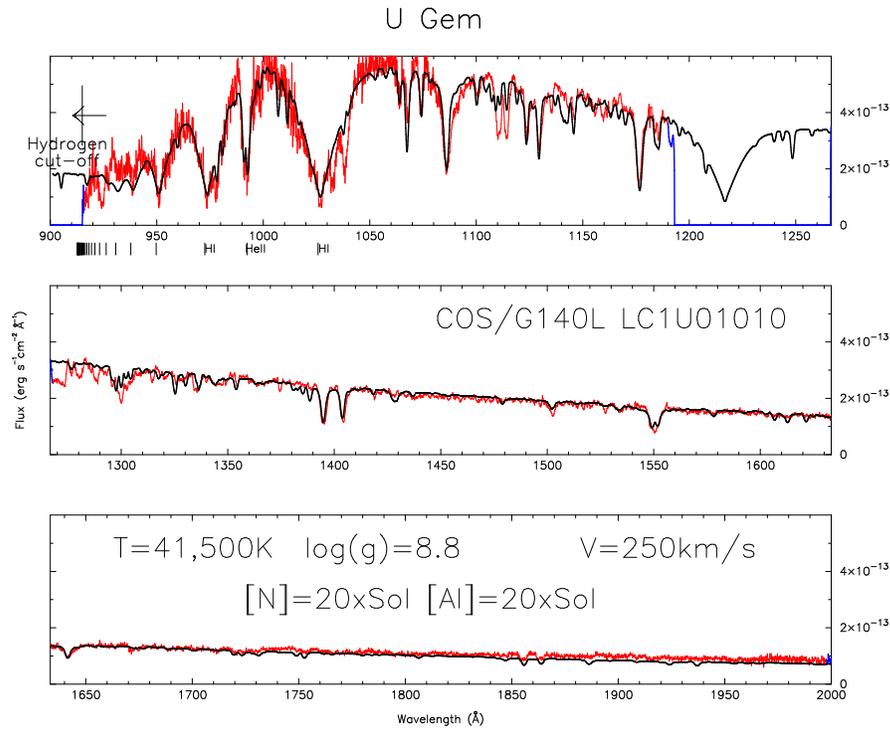}          
\caption{
The COS/G140L (center wavelength 1280~\AA ) spectrum of U Gem (in red)  
obtained at epoch \#1, 15 days after the peak of the outburst,
is modeled with a WD photosphere spectrum (in black). 
The observed spectrum extends from $\sim$915~\AA\ (the hydrogen
Lyman limit) to $\sim$2150~\AA\ ,  
while the theoretical spectrum extends from 900~\AA\ to 3200~\AA\ . 
The spectral regions of the observed spectrum 
that are matched to the theoretical spectrum
are in red, the excluded regions are
in blue; namely, $\lambda < 915$~\AA , $\lambda > 2000$~\AA , and
the gap betweens the two spectral segments overlaping the hydrogen 
Ly$\alpha$ region $\sim 1190-1268$~\AA .
The model has $\log(g)=8.8$ and is scaled to the distance of 
$\sim$100pc,  giving a 
a temperature of 41,500~K, 
with a projected rotational velocity of 250~km~s$^{-1}$. The abundances 
are all solar, except for nitrogen and aluminum which had to be 
set to 20 times solar to fit the absorption features.      
Note in the very long wavelengths, the model
is slightly deficient  in flux, which could be due to the small
contribution from a very weak component, but could also be due 
to some detector effect as shown in the lower panel of 
Fig.\ref{rawsubex}.
This spectrum is a combination of 12 sub-exposures and has the
best S/N, but consequently it also suffers from line broadening.
\label{twd1} 
} 
\end{figure*}

In order to match the profile of the Lyman series (down to 
$\sim 940$~\AA ), 
we had  to increase the temperature to 41,500~K, while
the continuum of the model in the longer wavelengths slightly decreased.
The Lyman series is much more sensitive to the temperature than the
continuum slope in the longer wavelength and therefore provides
a much more reliable way to determine the temperature of the WD.  
This model presents a satisfactory fit to the overall spectrum
and is presented in Fig.\ref{twd1}.  

Our grid of model for epoch 1 is spaced every 250~K, 
and we find that the best fit spans 3 models, from 
40,250~K to 40,750~K, giving limiting values of 40,000~K
and 41,000~K. We therefore take a 500~K error bar for 
the WD temperature, namely, 
for the first epoch spectrum we find a WD temperature of 41,500~K$\pm$500~K,  
15 days after the peak of the outburst.  
For comparison, a 40,500~K model has a flux that 
is 3\% too low at 1000~\AA\ when scaled from the flux at 1400~\AA\ ,
but the calibration errors of COS are 
$\sim$1.5\% at 1400~\AA\ . 

With a distance of 100.3~pc, the scaled WD radius  
is $R_{\rm wd}=$5,029~km,  corresponding to a 
WD mass of  $M_{\rm wd}=1.111~M_{\odot}$ at a temperature of 41,500~K
\citep{woo95,pan00}. The WD mass and radius give an 
effective surface gravity of 
$\log(g)=8.765$ which is consistent with our assumption
of $log(g)=8.8$. 
This model is a little deficient in flux at very short wavelengths
($< 950$~\AA ) and long wavelengths ($> 1725$~\AA\ ,  
see Fig.\ref{twd1}). A possible
explanation for the discrepancy in the very long wavelength region could
be a detector edges effects as apparent from Fig.\ref{rawsubex}.   

There are other possible emitting components in the system such as 
the hot spot (L1-stream impacting the rim of the disk), the boundary
layer, or the very low mass accretion rate disk which could 
contribute additional flux.  It has also been suggested that the 
WD itself could have a hot accretion belt remaining in its equatorial
region after ouburst, or the WD could have a non-uniform temperature
\citep{lon93,sio98,lon99,lon06}. 
As mentioned previously, not all the silicon lines 
can be fitted at the same time with the same abundance, however, this
could also be due to two emitting components with a different
temperature. For example the  Si\,{\sc ii} (1530) and 
Si\,{\sc iii} ($\sim1112$) lines are best fitted
at $\sim$30,000~K (assuming $log(g)=8.8$), while the other silicon
lines give a better fit at 35,000-45,000~K.  
We therefore tried two-temperature WD models,
with a hot equatorial region on the surface of a cooler WD. 

In order to fit the longer wavelength region, we started by taking 
a low WD temperature ($\sim 25,000$~K) model 
(with a corresponding larger radius)
that did not provide any flux in the short wavelength region, and added 
to it the contribution from a hotter region ($T \sim 45,000$~K and higher) 
to provide
flux in the shorter wavelength region. Such models did not fit
the COS spectrum better than the single WD models.
We varied the size and temperature of the hot equatorial region and the
temperature of the WD, but 
this combined model did not provide a qualitatively 
better fit to the single WD model and was not required by the data.
The number of two-temperature models that could fit the data was very large
(divergence of the solution). 
Therefore, in in our analysis below, we only considered single WD models.

The WD temperature depends on the assumed effective surface gravity 
($\log(g)$) of the model  as well as on the addition of a hot 
second component \citep{lon95,lon06}). 
The second component 
is not always detected  \citep{sio98}, either because of the
wavelength coverage (i.e. $\lambda > 1,000$~\AA ) or maybe 
simply because it is not always there. The nature of the second
component is still a matter of debate, as different physical models
for the second component fit the data just as well \citep{lon06}.  
Nevertheless, 
the WD temperature we obtained two weeks after the peak of the outburst
is in good agreement with previous
spectral analysis of the system in quiescence and decline  
\citep{sio98,lon99,fro01,lon06}.

\subsubsection{The Second Epoch.}

\begin{figure*}[t!] 
\vspace{-8.5cm} 
\plotone{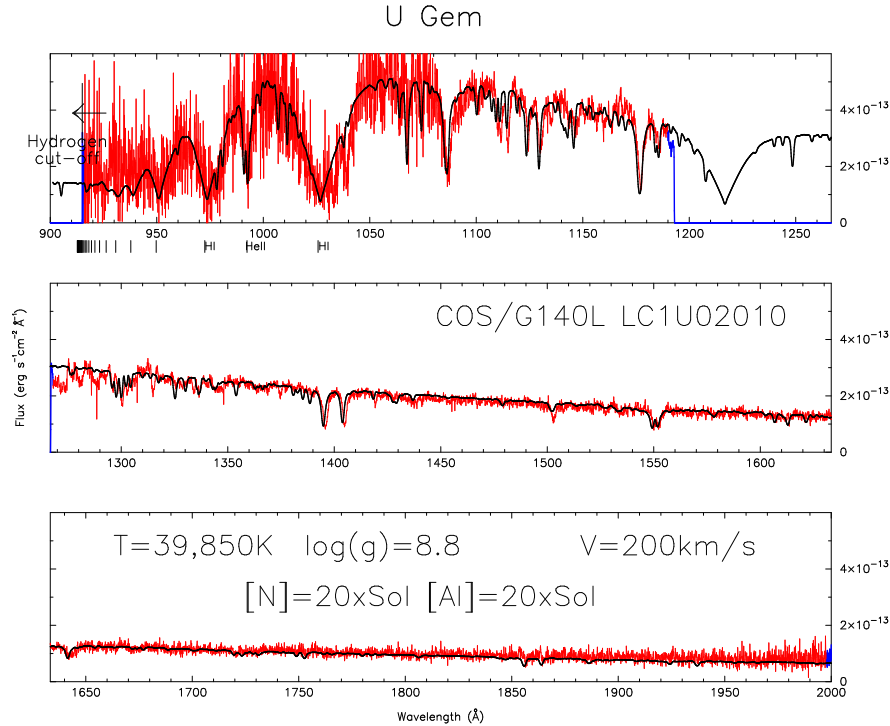}          
\caption{ 
The COS spectrum of U Gem   
obtained at epoch \#2, 21 days since the peak of the ouburst,
is modeled with a WD photosphere spectrum, as 
in the previous figure.  
The temperature obtained from the fitting is 39,850~K. 
The abundances 
are all  solar, except for nitrogen and aluminum which were  
set to 20 times solar.
This spectrum is pretty noisy, especially in the very short
wavelength region, as it is a combination of only 3 sub-exposures
(sub-exposure \#2 was lost due to the COS light path being blocked).  
\label{twd2} 
} 
\end{figure*} 

With only three sub-exposures, the second  epoch spectrum has a much
noisier spectrum, especially in the short wavelength region. 
Our single WD model fit gave a temperature of 39,850$\pm 500$~K , and a
scaled WD radius of 5,022~km, nearly equal to that obtained
for epoch \# 1. However, due to the slightly lower temperature,
this radius corresponds to a 1.109$M_{\odot}$ WD mass, or 
a $\log(g)$ of 8.766.  
This model fit is presented in Fig.\ref{twd2}. 
This spectrum was obtained 21 days after the peak of the outburst, 
exhibiting a cooling of 1,650~K in 6 days. 
The discrepancy in the long wavelength region is barely noticeable,
but it is still noticeable in the very short wavelength region. 
The broadening of the lines (200~km~s$^{-1}$) includes some 
orbital velocity broadening as well as lines forming in 
different regions, as the projected WD rotational
velocity is known to be about 100-150~km~s$^{-1}$.

\subsubsection{The Third Epoch.} 

\begin{figure*}[t!] 
\vspace{-8.5cm} 
\plotone{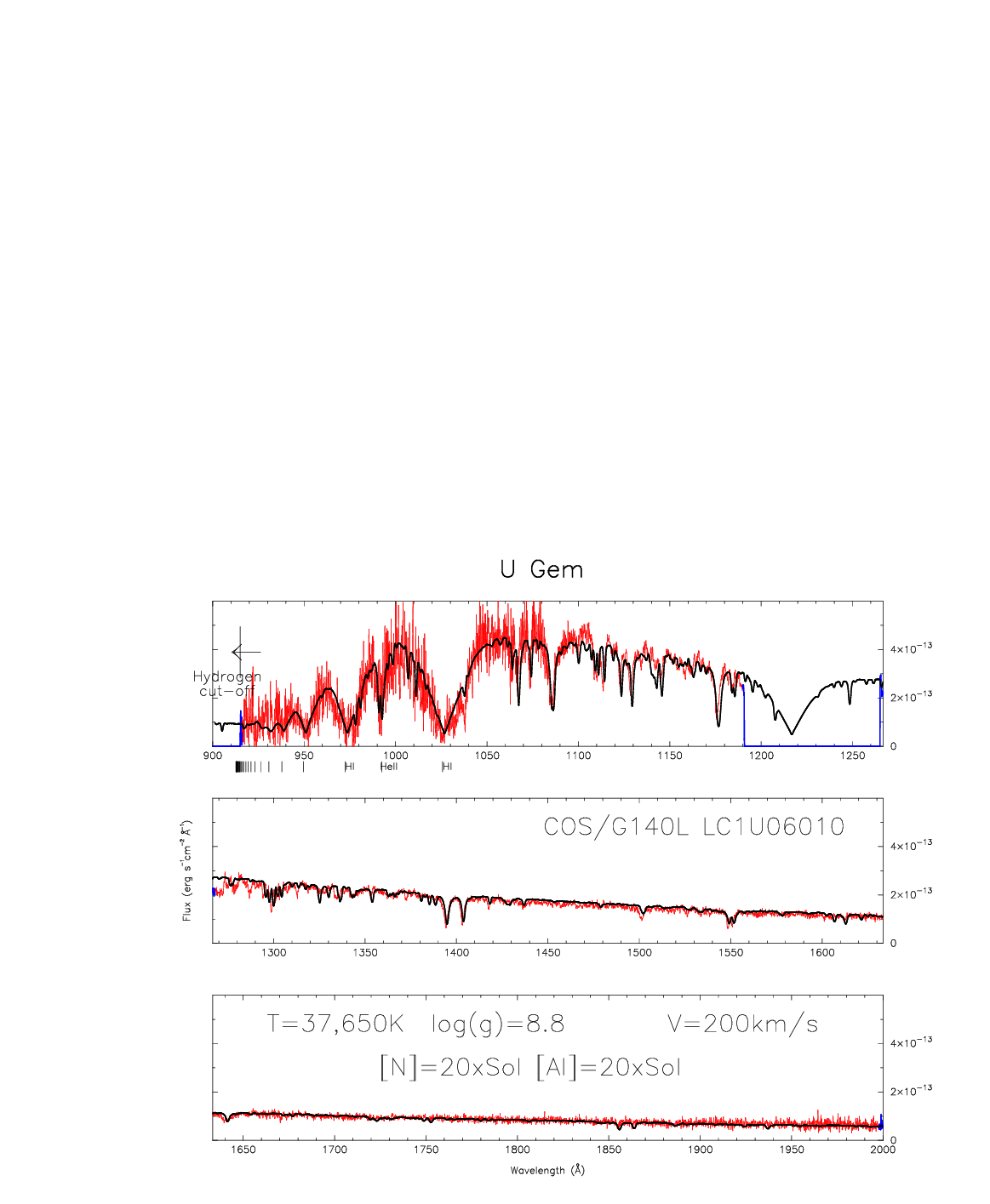}          
\caption{
The COS spectrum of U Gem obtained at epoch \#3, 33 days after the 
peak of the outburst, is modeled with a 37,650~K WD model. 
Except for the WD temperature, all the other parameters are the  
same as in the modeling carried for epoch \#2.
The spectrum was generated by co-adding the 4 sub-exposures 
obtained in one ({\it HST}) orbit.  
\label{twd3} 
} 
\vspace{-8.0cm} 
\plotone{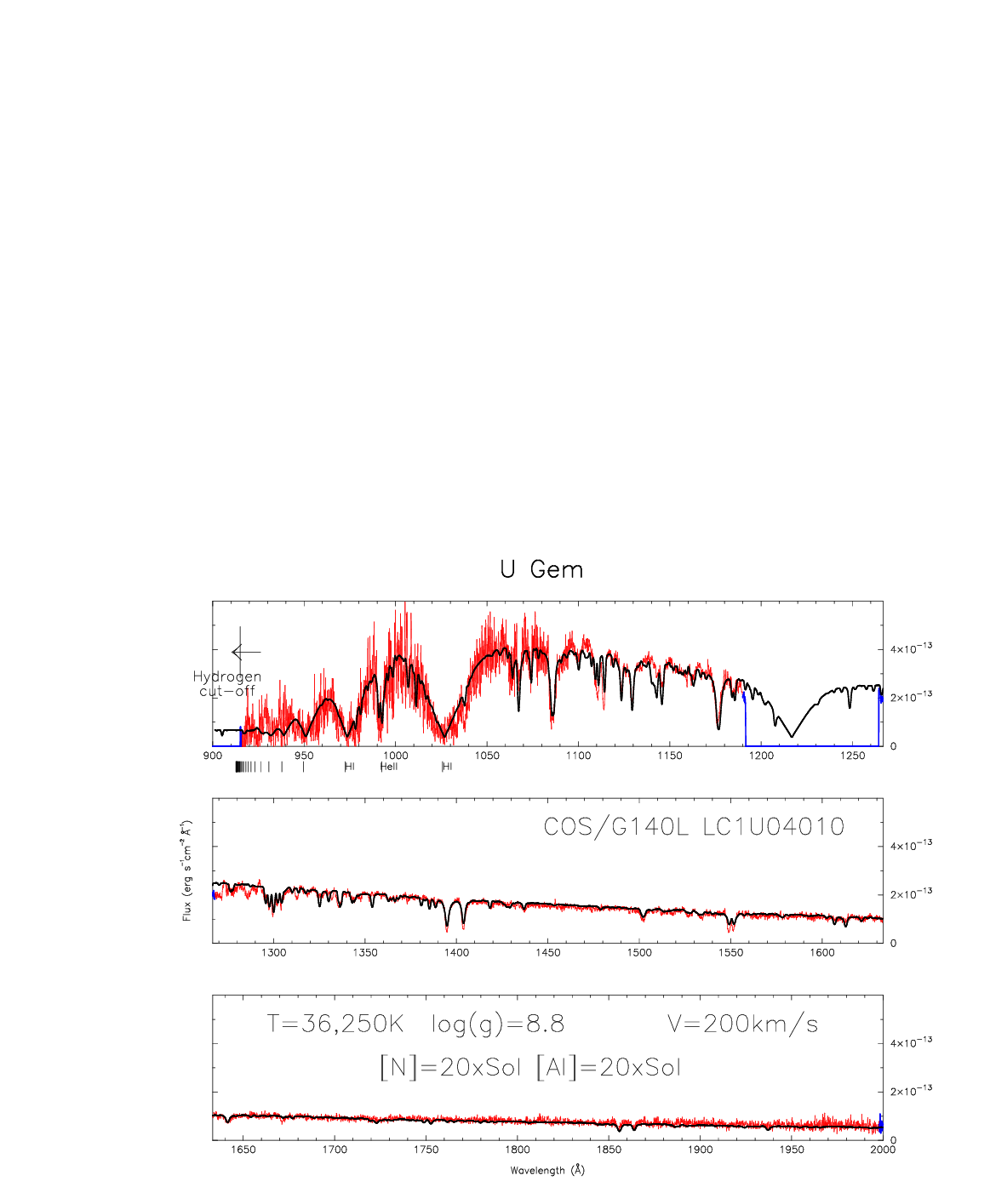}          
\caption{ 
The COS spectrum of U Gem obtained at epoch \#4, 56 days after the 
peak of the outburst, is modeled with a 36,250~K WD model. 
All the other parameters of the model are as specified  
for the modeling of epochs \#2 and \#3.         
The spectrum was generated by co-adding the 4 sub-exposures 
obtained in one ({\it HST}) orbit.  
\label{twd4} 
} 
\end{figure*}

With 4 sub-exposure spectra obtained 33 days after the 
outburst peak, the third epoch spectrum yielded 
a WD temperature of 37,650$\pm 500$~K and a scaled radius of 4,999~km.
This corresponds to a WD mass of $1.109M_{\odot}$ and an effective
surface gravity of $\log(g)=8.77$. 
In 17 days (since epoch \# 2), the WD temperature dropped by  
an additional 2200~K. 
Compared to the two previous epoch spectra, the model fit
(Fig.\ref{twd3}) has improved, there is no more discrepancy in the long 
wavelength region, and it is barely noticeable in the short 
wavelength region.

\subsubsection{The Fourth Epoch.} 

Also with 4 sub-exposure spectra, the fourth and last epoch spectrum
was fitted with a single WD model with a temperature of 36,250$\pm 500$~K  
(Fig.\ref{twd4}),  
revealing a further cooling of 1400~K in 23 days (i.e. 56 days
after the outburst peak). 
The scaled radius obtained was 5,005~km, 
corresponding to a WD mass of $1.107M_{\odot}$ and an effective
surface gravity of $\log(g)=8.768$. 
Here too, there is no more discrepancy in the long 
wavelength region, and very little discrepancy in the short 
wavelength region.

\begin{figure}[h] 
\vspace{-3.0cm} 
\plotone{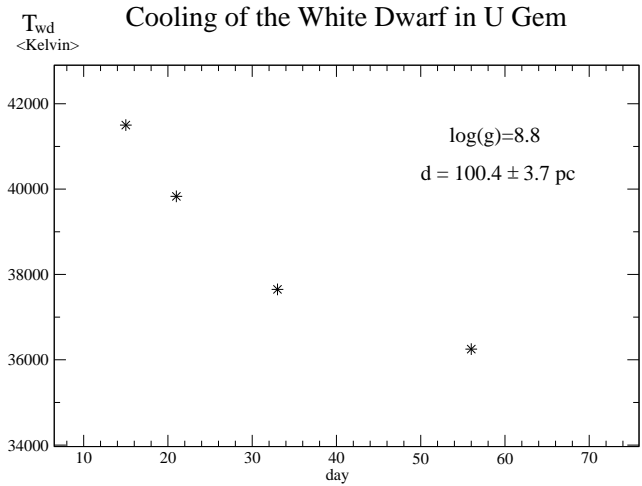}          
\caption{
The effective surface temperature of the WD is shown as a function 
of time since the peak of the outburst. The temperature were obtained
assuming a massive WD with $log(g)=8.8$ and a distance of $\sim100$ pc.
The ouburst was a wide outburst lasting about 15 days.  
\label{twdvstime} 
} 
\vspace{-0.8cm} 
\end{figure}

For the third and fourth epoch spectra, the fit to the continuum is
relatively good and there is little indication of a possible 
second emitting component. We find that our results are self-consistent
with a massive WD with an effective surface gravity of 
$\log(g)\approx 8.77$ (as there is basically
no difference between theoretical spectra with $\log(g)=8.77$ or $\log(g)=8.8$ 
- which was our initial assumption), and a radius of 
about 5,000~km. If we take the error in the distance, of about 4\%,   
into account, we find an error in the WD radius of 200~km, giving
$R_{\rm wd} = 5000\pm 200$~km. 
We find that the WD cools by 5,250~K in 41 days, 
from a temperature of 41,500~K 15 days after the peak
of the outburst, down to 36,250~K 56 days after the peak of the
outburst. The cooling of the WD is displayed in Fig.\ref{twdvstime}.   
The WD radius and temperatures we obtained in our analysis are consistent with
the results from the {\it FUSE} spectral analysis 
of \citet{lon06}, who obtained a temperature of 
$T_{\rm eff} \sim 43,600-47,100$ (for $\log(g)\sim 8.5-9.0$, respectively) 
10 days after the peak of a wide outburst, and 
$T_{\rm eff} \sim 30,300-31,600$ (for $\log(g)\sim 8.5-9.0$) 
135 days after the peak of a (different but similar) wide outburst.

\section{{\bf Summary and Conclusion}} 

We have obtained {\it HST}/COS spectra of U Gem following
a wide outburst when the WD dominated the FUV. 
The spectra were taken at 4 epochs, 15 days, 21 days, 33 days,
and 56 days  after the peak of the outburst, and reveal 
a  decreasing FUV flux as a function of time, an obvious sign
of the cooling of the WD.

\begin{enumerate}

\item
The {\it HST}/COS spectra of U Gem exhibits the usual hydrogen Lyman 
and helium Balmer absoprtion lines forming in the WD photosphere, 
as well as absorption lines of 
low ionization energy species of carbon (C\,{\sc ii iii iv}), 
nitrogen (N\,{\sc iii}), sulphur (S\,{\sc iv}) and silicon 
(Si\,{\sc ii iii iv}).  We also identified absorption 
lines of iron (Fe\,{\sc ii}), aluminum
(Al\,{\sc iii}),  and possibly also argon (Ar\.{\sc iii}).
All the metal lines  could potentially form in the photosphere of a moderately
hot WD. However, many of them (i.e. Si, Al, Ar) are deeper at orbital phases
$\phi \sim 0.25$ and $\phi=$0.67-0.81, indicating the presence of 
absorbing material at these specific phases.  
Higher ionization energy species  of N\,{\sc iv} and O\,{\sc vi} 
seen in absorption are known to form at a much higher temperature
($\sim 80,000$~K) and are not attributed to the WD.  
The spectra do not show any emission features.

\item 
The first epoch spectra cover
en entire orbital phase and show a variation in the continuum flux level 
with the orbital phase that is more pronounced in the short 
wavelength range. The FUV light curves display an increase in flux near 
phase $\phi=0.9$, a small dip near phase $\phi=0.25$ and a stronger
dip near phase $\phi=0.78$. The increase in flux at $\phi=0.9$ is  
most likely due to the hot spot facing the observer \citep{und06,ech07}. 
The dip around
phase $\phi=0.78$ is due to L1-stream material overflowing
the disk rim and falling toward smaller radii, where it has been
shown to bounce back off the disk plane near phase $\phi=0.5$;    
there it is re-directed toward $\phi \approx 0.2$ \citep{kun01}, where 
it is likely responsible for the small dip observed in the FUV lightcurve
at $\phi=0.25$.   
The overflowing L1-stream material is also responsible for the 
strong orbital 
phase variation of the depth of the the absorption lines of 
silicon, aluminum and argon.   

\item 
Our analysis of absorption lines velocity as a function of the
binary orbital phase reveals an overall red-shift in the velocity
of the silicon lines, especially Si\,{\sc iii} \& {\sc iv},
compared to the WD orbital velocity, indicating material falling
towards the WD. 
The largest velocity shift in the lines is observed around phase
$\phi \sim 0.78-0.91$.  
The carbon lines appear to follow more closely the
WD orbital velocity, implying that they are probably forming in the
WD photosphere.

\item
The spectra exhibit solar to sub-solar abundances of carbon and sulfur, 
supra-solar (up to $20 \times$) abundances of nitrogen, aluminum, and argon, 
while iron is only marginally supra-solar ($2 \times$).  
Silicon is sub-solar except for the Si\,{\sc iii} ($\sim 1112$~\AA ) and
Si\,{\sc iv} (1400~\AA ) that are solar.
Our analysis indicates that the silcon, aluminum and
argon lines are at least in part forming in material above the disk
veiling the WD at orbital phases $\phi \sim 0.25$ and $0.75$.    
In spite of the fact that many lines might be affected by absorbing
material in front of the WD at all orbital phases,  
our analysis  confirms the high 
N/C ratio observed previously in the FUV spectra of U Gem in quiescence, 
a sign of possible CNO enrichment of the accreted material on the 
WD surface. 

\item
At the time of the first epoch, 15 days after the peak of the outburst,
the WD was still rather hot reaching a temperature of 41,500~K.
By the time of the 4th epoch, 41 days later, the WD had cooled down to 
a temperature of 36,250~K. 
The temperatures we obtained are consistent with previous
observation of U Gem in quiescence for the $\log(g)$ considered here
\citep{lon06} and are theoretically consistent with compressional heating
and subsequent cooling of the WD as shown by \citet{sio95}.    

\item 
The spectra are consistent with  a single temperature WD component, 
and two-temperature  WD models did not provide any improvment in the fit.  
Our single WD temperature model fits are self-consistent with a massive WD 
with an effective surface gravity $\log(g)=8.77$ and 
a ($\sim 40,000$~K) WD radius of $5000\pm 200$~km corresponding  
to a WD mass $M_{\rm wd} \approx 1.11 M_{\odot}$ and consistent with
previous spectral analysis of U Gem in quiescence \citep{lon06}.

\item 
We confirm that phase-dependent absorption is the most plausible
interpretation of the time-variable absorption seen in the FUV  
spectra of U Gem 
during quiescence, and we concur with \citet{lon06} that it  
complicates the analysis of the WD spectra. 

The picture that emerges from the orbital phase variability of the spectra
is a familiar one. 
The L1-stream material hits the disk's rim near $\phi=0.81-0.91$ creating 
a peak in the UV light curve at that orbital phase. Some of the material
overflows the disk's edge and moves to smaller radii where it reduces 
the UV flux at phase 0.78, increases the depth of some absorption lines
at $\phi = 0.78-0.81$, and adds a red-shifted component to absorption
lines near $\phi=0.78-0.91$. The overflowing material eventually falls
back onto the disk 
near phase 0.5 (as corroborated by Doppler mapping \citet{ech07} 
and simulations \citet{kun01}) where it bounces back off the disk 
and continues in 
a trajectory toward $\phi=0.25$. At this phase the material is high enough
above the disk to veil the 
WD for a second time, creating 
a second dip in the UV lightcurve as well deeper absorption lines
in the spectrum.

\end{enumerate}

\vspace{-0.8cm} 
\acknowledgements 
 
PG is pleased to thank William (Bill) P. Blair at the Henry Augustus Rowland 
Department of Physics and Astronomy at the Johns Hopkins University, 
Baltimore, Maryland, USA, for his kind hospitality. 
We wish to thank an anonymous referee for her/his comments that helped improve
the manuscript. 
We have used some of the online data from the AAVSO, and are thankful to the 
AAVSO and its members worldwide for making this data public and for their 
constant monitoring of cataclysmic variables.   
This research was based on observations made with the NASA/ESA 
Hubble Space Telescope, obtained at the Space Telescope Science Institute,
located in Baltimore, Maryland, USA. 
Support for program \#12926 was provided by NASA throught a grant from the
Space Telescope Science Institute, which is operated by the Association of
Universities for Research in Astronomy, Inc., under NASA contract NAS 5-26555. 

\facilities{{\it HST}(COS)} 

\software{IRAF, FORTRAN, PGPLOT, TLUSTY / SYNSPEC, all installed and running
under Cygwin-X}


\vspace{-0.2cm} 

\begin{thebibliography} 

\vspace{-0.6cm} 

\bibitem[Bruch \& Engel(1994)]{bru94}
Bruch, A., \& Engel, A. 1994, \aaps, 104, 79  

\bibitem[Dai \& Qian(2009)]{dai09}
Dai, Z., Qian, S. 2009, ApSS, 321, 91 

\bibitem[Echevarr\'ia et al.(2007)]{ech07}
Ecchevarr\'ia, J., de la Fuente, E., Costero, R. 2007, \apj, 134, 262 

\bibitem[Froning et al.(2001)]{fro01} 
Froning, C.S., Long, K.S., Drew, J.E., Knigge, C., Proga, D. 2001, \apj, 
562, 963

\bibitem[Godon et al.(2012)]{god12}
Godon, P., Sion, E.M., Levay, K. et al. 2012, \apjs, 203, 29  

\bibitem[Hamada \& Salpeter(1961)]{ham61}
Hamada, T., \& Salpeter, E.E. 1961, \apj, 134, 683 

\bibitem[Harrison et al.(2004)]{har04} 
Harrison, T.E., Johnson, J.J., MaArthur, B.E., Benedict, G.F., 
Szkody, P., Howell, S.B., \& Gelino, D.M. 2004, \aj, 127, 460 

\bibitem[Hind(1856)]{hin56}
Hind, J.R. 1856, \mnras, 16, 56 

\bibitem[Hubeny(1988)]{hub88} 
Hubeny, I. 1988, Comput.Phys.Commun., 52, 103 


\bibitem[Hubeny \& Lanz(1995)]{hub95} 
Hubeny, I., \& Lanz, T. 1995, \apj, 439, 875  

\bibitem[Hubeny \& Lanz(2017a)]{hub17a} 
Hubeny, I., \& Lanz, T. 2017a, 
{\it A brief introductory guide to TLUSTY and SYNSPEC} \\
arXiv:1706.01859 

\bibitem[Hubeny \& Lanz(2017b)]{hub17b} 
Hubeny, I., \& Lanz, T. 2017b, 
{\it TLUSTY User's Guide II: Reference Manual} \\
arXiv:1706.01935 

\bibitem[Hubeny \& Lanz(2017c)]{hub17c} 
Hubeny, I., \& Lanz, T. 2017c, 
{\it TLUSTY User's Guide III: Operational Manual} \\
arXiv:1706.01937 

\bibitem[Kiplinger et al.(1991)]{kip91} 
Kiplinger, A.L., Sion, E.M., \& Szkody, P. 1991, \apj, 366, 569 

\bibitem[Kraft(1962)]{kra62}
Kraft, R.P. 1962, \apj, 135, 408 

\bibitem[Krzeminski(1965)]{krz65}
Krzeminski, W. 1965, \apj, 142, 1051 

\bibitem[Kunze et al.(2001)]{kun01} 
Kunze, S., Speith, R., \& Hessman, F. 2001, \mnras, 322, 499   

\bibitem[la Dous(1991)]{lad91}
la Dous, C. 1991, \aap, 252, 100  

\bibitem[Long \& Gilliland(1999)]{lon99} 
Long, K.S., Gilliland, R.L. 1999, \apj, 511, 916 

\bibitem[Long et al.(1993)]{lon93}
Long, K.S., Blair, W.P., Bowers, C.W., Davidsen, A.F., Kriss, G.A.,
Sion, E.M., \& Hubeny, I. 1993, \apj, 402, 327 

\bibitem[Long et al.(1995)]{lon95}
Long, K.S., Blair, W.P., Raymond, J.C. 1995, \apj, 454, L39 

\bibitem[Long et al.(2006)]{lon06}
Long, K.S., Brammer, G., Froning, C.S. 2006, \apj, 648, 558 

\bibitem[Long et al.(1996)]{lon96} 
Long, K.S., Mauche, C.W., Raymond, J.C., Szkody, P., \& Mattei, J.A. 
1996, \apj, 469, 841 

\bibitem[Long et al.(1994)]{lon94}
Long, K.S., Sion, E.M., Huang, M., \& Szkody, P. 1994, \apj, 424, L49 


\bibitem[Marsh et al.(1990)]{mar90}
Marsh, T.R., Horne, K., Schlegel, E.M., Honeycutt, R.K., 
Kaitchuck, R.H. 1990, \apj, 364, 637 

\bibitem[Mason et al.(1988)]{mas88}
Mason, K.O., C\'ordova, F.A., Watson, M.G., \& King, A.R. 1988, \mnras, 
232, 779 


\bibitem[Panei et al.(2000)]{pan00}
Panei, J.A., Althaus, L.G., \& Benvenuto, O.G. 2000, \aap, 353, 970 

\bibitem[Panek \& Eaton(1982)]{pan82}
Panek, R.J., \& Eaton, J.E. 1982, \apj, 258, 572 

\bibitem[Panek \& Holm(1984)]{pan84}
Panek, R.J., \& Hom, A.V. 1984, \apj, 277, 700 


\bibitem[Ritter \& Kolb(2003)]{rit03}
Ritter, H., \& Kolb, U. 2003, \aap, 404, 301 
(update RKcat7.24, 2016) 

\bibitem[Sion(1995)]{sio95}
Sion, E.M. 1995, \apj, 438, 876 

\bibitem[Sion et al.(1997)]{sio97}
Sion, E.M., Cheng, F.H., Szkody, P., Huang,M., Provencal, J.
Sparks, W., Abbott, B., Hubeny, I. 
Mattei, J., \& Shipman, H. 1997, \apj, 483, 907 

\bibitem[Sion et al.(1998)]{sio98}
Sion, E.M., Cheng, F.H., Szkody, P., Sparks, W., G\"ansicke, B.T., 
Huang, M., \& Mattei, J. 1998, \apj, 496, 449 

\bibitem[Smak(1971)]{sma71}
Smak, J.I. 1971, {\it Acta Astron.}, 21, 15  

\bibitem[Smak(1976)]{sma76}
Smak, J.I. 2976, {\it Acta Astron.}, 26, 277 

\bibitem[Smak(2001)]{sma01}
Smak, J.I. 2001, {\it Acta Astron.}, 51, 279 


\bibitem[Szkody and Mattei(1984)]{szk84}
Szkody, P., and Mattei, J. 1984, \pasp, 96, 988 

\bibitem[Szkody et al.(1996)]{szk96} 
Szkody, P., Long, K.S., Sion, E.M., \& Raymond, J.C. 1996, \apj, 469, 834 

\bibitem[Unda-Sanzana et al.(2006)]{und06}
Unda-Sanzana, E., Marsh, T.R., and Morales-Rueda, L. 2006, \mnras, 369, 805 

\bibitem[Verbunt(1987)]{ver87}
Verbunt, F. 1987, \aaps, 81, 339 

\bibitem[Wade(1981)]{wad81}
Wade, R.A. 1981, \apj, 246, 215 


\bibitem[Warner and Nather(1971)]{war71} 
Warner, B., and Nather, R.E. 1971, \mnras, 152, 219 

\bibitem[Wood(1995)]{woo95}
Wood, M.A. 1995, in White Dwarfs, Proc. 9th Europ.Workshop on WDs,
eds. D. Koester \& K. Werner (Lecture Notes in Physics, Vol.443; 
Berlin: Springer), 41 

\bibitem[Zhang and Robinson(1987)]{zha87}
Zhang, E.H., \& Robinson, E.L. 1987, \apj, 321, 813  

\end{thebibliography}
\end{document}